\documentclass[twocolumn,superscriptaddress,floatfix,longbibliography,aps,pra,nopreprintnumbers,noeprint,10pt]{revtex4-2}

\usepackage[utf8]{inputenc}
\usepackage[english]{babel}
\usepackage{graphicx}
\usepackage{bm}
\usepackage{bbm}
\usepackage{xcolor}
\usepackage{amsmath}
\usepackage{amssymb}
\usepackage{epstopdf}
\usepackage[normalem]{ulem} 
\usepackage{enumerate}
\usepackage[urlcolor=blue,colorlinks=true,citecolor=blue,linkcolor=blue,pdfstartview={FitH},bookmarks=false]{hyperref}
\renewcommand{\vec}{\bm}

\graphicspath{{Figures/}{./Figures/}{.}}

\begin{document}

\title{On the Relation Between String Order Parameters, Entanglement, and Dynamical Quantum Phase Transitions in Topological Dynamics}

\author{Sirshendu Bhattacharyya}
\affiliation{Department of Physics, Raja Rammohun Roy Mahavidyalaya, Radhanagar, Hooghly 712406, India}
\author{Szczepan G{\l}odzik}
\affiliation{Institute of Physics, Maria Curie-Sk\l{}odowska University, 20-031 Lublin, Poland}
\author{Nicholas Sedlmayr}
\email[e-mail: ]{sedlmayr@umcs.pl}
\affiliation{Institute of Physics, Maria Curie-Sk\l{}odowska University, 20-031 Lublin, Poland}

\date{\today}

\begin{abstract}
Topological order is defined by topological invariants, rather than symmetries and local order parameters. Nonetheless some topological phases can be characterized by string order parameters and entanglement. In this article we study how string order parameters and entanglement spectra behave out-of-equilibrium following quenches in one dimensional topological models with $\mathbb{Z}$ invariants. Previously it has been suggested that string order parameters could serve as an experimental probe of dynamical quantum phase transitions. Despite the existence of clear zeroes in the order parameters at critical times, we show that in general there is no exact quantitative or qualitative connection with the critical times of dynamical quantum phase transitions. Another possible connection is of dynamical string order parameter zeroes and dynamical crossings at the center of entanglement spectra. Here we see that there can sometimes be a connection, but it is not typical. Again there is no general quantitative or qualitative connection. Each dynamical form of criticality behaves independently, though we do see that critical times tend to be of the same order of magnitude and give an argument for why this is the case. We also find that a string order parameter which labels one topological phase can undergo non-trivial dynamics even following a quench between \emph{other} topological phases. We elucidate where connections can be made, and where they result from a consideration of insufficiently general models. These results cast doubt on the idea of genuine dynamical phases following quenches in such models.
\end{abstract}

\maketitle

\section{Introduction}\label{sec:intro}

Topological order refers to phases which are characterized by a topological invariant rather than a local order parameter or by symmetries~\cite{Ryu2010,Hasan2010,Trifunovic2021}. Physically one can think of a gapped band structure as possessing a topology. A topological
phase transition occurs when the gap closes allowing the invariant to change. Due not only to the difficulty in directly measuring a topological invariant experimentally, but also as it is not always possible to calculate an invariant, or to know which invariant to calculate, alternative proxies for the topology are desired. Some are based on the justly famed bulk-boundary correspondence~\cite{Chiu2016,Trifunovic2018}, allowing one to investigate the existence of edge modes rather than the bulk topology directly~\cite{Glodzik2023}. Alternatives include using entanglement measures and also for certain one dimensional models non-local ``string order parameters '' can be defined for specific topological phases, which replace local order parameters. String order parameters were first identified in spin models such as the Haldane model~\cite{DenNijs1989,Hida1992,Hida1992a} and later were applied to symmetry protected topology~\cite{Chitov2018,Bahovadinov2019,Sorensen2023}. Generalizations to higher winding numbers are possible, but non-trivial, and connections to multi-partite entanglement have also been discussed~\cite{Pezze2017,Zhang2018}.

It has been found that string order can disappear after even infinitesimal times following a quench. In this case an integer spin Heisenberg chain was prepared in the Affleck-Kennedy-Lieb-Tasaki state. Following a symmetry breaking quench string order disappeared suddenly~\cite{CalvaneseStrinati2016}. String order was also found to be unstable to small perturbations in equilibrium~\cite{Anfuso2007}. It would perhaps therefore be surprising if it turned out to be a robust indicator of dynamical topological properties of a system.

A dynamical quantum phase transition (DQPT) is the name given to a non-analyticity which occurs for a dynamical analogue of the free energy~\cite{Heyl2013,Andraschko2014,Heyl2018a,Heyl2019,Sedlmayr2019a}. This analogue is the return rate, defined in terms of the Loschmidt amplitude, \emph{i.e.}~the overlap between a time evolved and initial state, and the non-analyticity at critical times is related to zeroes of the Loschmidt amplitude~\cite{Heyl2013,Schmitt2015,Vajna2015,Maslowski2024b} and can be measured in ion traps and cold atom settings~\cite{Jurcevic2017,Flaschner2018,Zhang2017b,Guo2019,Smale2019,Nie2020,Tian2020}. DQPTs were also studied in topologically non-trivial systems~\cite{Schmitt2015,Vajna2015,Jafari2016,Bhattacharya2017,Jafari2017a,Sedlmayr2018,Jafari2018,Zache2019,Maslowski2020,Okugawa2021,Rossi2022,Maslowski2023,Maslowski2024,Maslowski2026a}, and dynamical order parameters, which count the zeroes of the Loschmidt amplitude have been introduced~\cite{Budich2016,Heyl2017,Bhattacharya2017a}. In general there is however no sense of a well defined dynamical phase separated by DQPTs~\cite{Cheraghi2023}. Dynamically each physical quantity may need to be considered separately. A possible example of a kind of dynamical phase is of a dynamical form of topology of the Loschmidt amplitude which gives rise to a dynamical bulk boundary effect, with Loschmidt zeroes formed between critical times~\cite{Sedlmayr2018,Maslowski2020,Maslowski2023,Maslowski2024,Maslowski2026a,Maslowski2026b}.

Due to the difficulty of measuring the Loschmidt echo, connections between DQPTs and observables such as order parameters are desired. Connections have been proposed between critical times of DQPTs and either zeroes or periodic modulation of order parameters~\cite{Heyl2013,Karrasch2013,Heyl2014,Wrzesniewski2022}. Analogously in one dimensional topological insulators and superconductors the string order parameter has been suggested to have a connection with DQPTs, becoming zero at the critical times, which has been demonstrated numerically in a number of models~\cite{Budich2016,Halimeh2018,Hagymasi2019,Uhrich2020}. It has also been suggested that entanglement, as measured by entanglement entropy~\cite{Bennett1996,Vidal2003a}, may have connections to DQPTs~\cite{Torlai2014,Sedlmayr2018,Maslowski2020,Poyhonen2021}, though this can not be a direct one-to-one relationship~\cite{Gong2018a}. We check also the relationship of entanglement to the string order parameters.

In equilibrium one finds that a (topological) quantum phase transition cleanly divides different behaviors for a variety of physical phenomena. Finite size corrections however may have their transitions at slightly different values of the control parameter. In this article we argue that for dynamical transitions the situation is more complicated. Even in the thermodynamic limit different measures of the phase or behavior of the system may have different critical times. This further casts doubt on the idea of \emph{general} topological dynamical phases, or connections between different types of dynamical criticality, string order, or entanglement.

In this work we on the one hand want to demonstrate independence between different dynamical properties. On the other hand we will also argue that behind different kinds of criticality lies broadly similar physics, which explains why these phenomena are often seen together, and can have similar critical times. Although no general connection to DQPTs is found, we do see occasional connections between string order parameter zeroes and entanglement spectra crossings, though these are not generic phenomena. We also find that string order parameters are generally not able to distinguish broadly different quench scenarios, \emph{i.e.}~quenches between different topological phases.

This article is organized as follows. In sections \ref{sec:sop} and \ref{lrkc} we introduce the two exemplary models we will focus on, calculate the string order parameters, and explain the equilibrium behavior. In section \ref{sec:ee} we introduce the other dynamical measures we compare the string order parameters to. In \ref{sec:results} and \ref{sec:kitres} we present a selection of results for the dynamics of the two models following different quench protocols, and compare the critical behavior of the string order parameters, entanglement spectra, and DQPTs. In section \ref{sec:conc} we discuss the results and conclude. Throughout this article we scale $\hbar=1$.

\section{String order parameter for the SSH Model}\label{sec:sop}

\subsection{The Model}\label{sec:sshmodel}

In this article we focus on two different models. First we use the widely studied Su-Schrieffer-Heeger (SSH) model~\cite{Su1980,Sirker2014}. The relative simplicity of this model allows for some simplified expressions to be found, as well as highly accurate numerics. However, as we will see, its very simplicity also tends to mislead with some of its simple results which do not generalize to even the simplest extension. In the next section \ref{sec:kitmodel} we then use a generalized Kitaev chain, with longer range hopping and p-wave pairing terms included to reach a wider range of winding numbers. In each of these cases we derive the string order parameter independently. Although a unified description is possible, this tends to obscure some of the simplicity of each case.

The SSH model can be thought of as a chain of $N$ unit cells with each unit cell consisting of two different sites: $a$ and $b$. There is only a hopping term present which alternates with strengths $-J(1\pm\delta)$. The Hamiltonian of this chain is defined by
\begin{equation}
H = -J\sum\limits_{i=1}^{N}\left[  (1-\delta)a^{\dagger}_i b_i + (1+\delta) a^{\dagger}_{i+1} b_i \right]+\textrm{H.c.}\,,
\label{ssh}
\end{equation}
where $a_i,\;b_i$ are fermionic annihilation operators and $N$ is the number of unit cells. To diagonalize the Hamiltonian we use the Fourier transformation
\begin{align}
a_j =& \frac{1}{\sqrt{N}}\sum\limits_{k}e^{-ikj}a_k\,, \nonumber \\
b_j =& \frac{1}{\sqrt{N}}\sum\limits_{k}e^{-ikj}b_k\,.
\end{align}
where $a_k$ and $b_k$ are respective fermionic annihilation operators on quantized momentum $k$. The transformation results in
\begin{equation}
H = \sum\limits_{k} \Gamma_k^{\dagger} \mathcal{H}_k \Gamma_k\,,
\label{ssh2}
\end{equation}
Here $H$ is scaled by $J$, $\Gamma_k^{\dagger} \equiv (a_k^{\dagger} \;\; b_k^{\dagger})$ and
\begin{equation}
\mathcal{H}_k =
\begin{pmatrix}
0 & 1-\delta+e^{ik}(1+\delta) \\
1-\delta+e^{-ik}(1+\delta) & 0
\end{pmatrix}\,.
\label{ssh-hk} 
\end{equation}
Diagonalizing $H_k$, we obtain the eigenvalues for each $k$ as
\begin{equation}
\varepsilon_k=\mp \sqrt{2}\sqrt{1+\delta^2 + (1-\delta^2)\cos k}
\end{equation}
with the corresponding eigenstates
\begin{equation}
|\psi_{k}\rangle_{\mp} = 
\begin{pmatrix}
\mp e^{i\phi_k} \\
1
\end{pmatrix}\,,
\end{equation}
with $\tan \phi_k = \frac{(1+\delta)\sin k}{1-\delta+(1+\delta)\cos k}$. The topological phase transition is marked at $\delta=0$. See appendix \ref{app:top} for information about the topological invariant $\nu$.

\subsection{The String Order Parameter}\label{sec:sshsop}

For full clarity we will proceed step by step in deriving first the equilibrium, and then the dynamical, string order parameter for the SSH model. Here we will give the main details of the string order parameters, further details can be found in appendix \ref{app:ssh}. The string order parameters, which can be found from the local order parameters of a spin chain following a Jordan-Wigner transformation, are defined as~\cite{Budich2016,Uhrich2020}
\begin{align}
\mathcal{O}^x_{l,m} =& \langle B^-_{l} A^+_{l+1} B^-_{l+1} \cdots A^+_{m} \rangle_0 \nonumber \\
\mathcal{O}^y_{l,m} =& \langle B^+_{l} A^-_{l+1} B^+_{l+1} \cdots A^-_{m} \rangle_0
\end{align}
where
\begin{equation}
A^\pm_j = (a^\dagger_j \pm a_j)\textrm{ and }
B^\pm_j = (b^\dagger_j \pm b_j)\,.
\end{equation}
Here we will focus on $\mathcal{O}^x$ which characterizes the topologically non-trivial phase $\nu=1$. The calculation for $\mathcal{O}^y$ follows similarly. We note that for each topological phase for each model under consideration a string order parameter needs to be found independently, for some other examples see Ref.~\onlinecite{Pezze2017,Zhang2018}. In all cases, the average is over the ground state $\Psi_0$ of some Hamiltonian $H_0$. When considering the dynamics this will be our initial state.

For an average over the ground state, Wick's theorem enables us to write $\mathcal{O}^x_{l,m}$ in the form of a Toeplitz determinant \cite{Lieb1961,Barouch1971}:
\begin{equation}
\mathcal{O}^x_{l,m} =
\begin{vmatrix}
G_{l,l+1} & G_{l,l+2} & \cdots & G_{l,m} \\
G_{l+1,l+1} & \cdots & \cdots & G_{l+1,m} \\
\vdots & & & \vdots \\
G_{m-1,l+1} & \cdots & \cdots & G_{m-1,m}
\end{vmatrix}
\end{equation}
with $G_{l,m}=\langle B^-_l A^+_m \rangle_0$ given by
\begin{equation}
G_{l,m} = -\frac{1}{N}\sum_k
\frac{(1+\delta)\cos[(m-l-1)k]+(1-\delta)}{\varepsilon_k}
\label{glm-sum}
\end{equation}
where $k=2\pi n/N$ and $n\in\{-N/2+1,\cdots,N/2$\}.

\begin{figure}
\includegraphics[width=0.95\columnwidth]{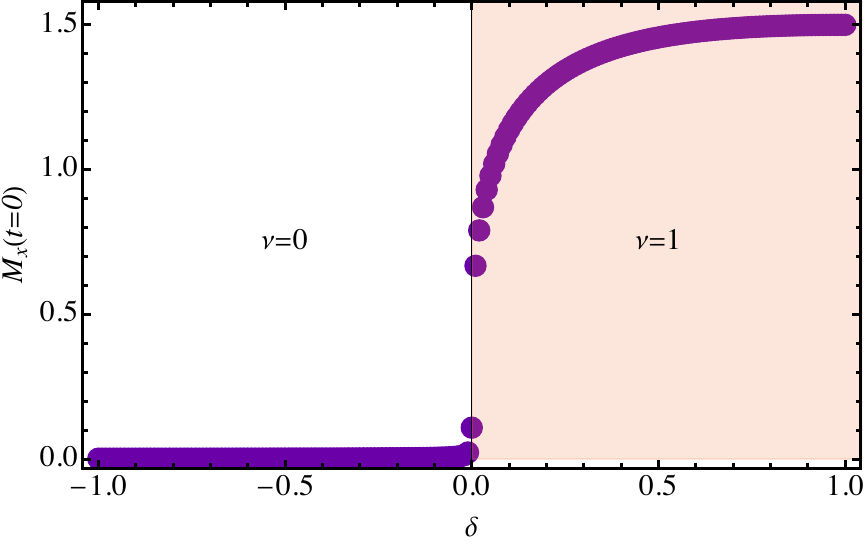}
\caption{Plots of $\mathcal{M}$ vs $\delta$ for the SSH model in equilibrium for $N=10^3$. One can see that in the topologically trivial phase, where $\nu=0$,  $\mathcal{M}=0$. For the topologically non-trivial phase $\nu=1$ we find $\mathcal{M}\neq0$.}
\label{fig:ssheq}
\end{figure}

In general $\mathcal{O}^x_{l,m}$ has some dependence on both the distance $m-l$ and the system size $N$. As we are interested largely in the bulk properties in the thermodynamic limit we therefore set $m-l=N/2-1$ and define
\begin{equation}
\mathcal{M}=\left|\mathcal{O}^x_{1,\frac{N}{2}}\right|\,.
\end{equation}
In Fig.~\ref{fig:ssheq} we plot $\mathcal{M}$ against $\delta$ for $N=10^3$, resulting in $\mathcal{M}(\delta>0)\neq0$ and $\mathcal{M}(\delta<0)\to0$. For $N=10^3$ the string order parameter has already converged.

Now we consider a sudden quench in the parameter $\delta_i\to\delta_f$ at time $t=0$, before which the system was assumed to be in the ground state of the Hamiltonian with dimerization $\delta_i$. Following the quench the string order parameter at any time $t$ can be written as~\cite{Uhrich2020}
\begin{equation}
\mathcal{O}^x_{lm}(t) =\left\langle B^-_{l}(t)\left[\prod\limits_{j=l+1}^{m-1} A^+_{j}(t)B^-_{j}(t)\right]A^+_{m}(t)\right\rangle_0
\label{Oxt}
\end{equation}
where $B^-_{i}(t)\equiv b^{\dagger}_i (t) - b_i (t)$, $A^+_{i}(t)\equiv a^{\dagger}_i (t) + a_i (t)$ and the time evolution is generated by the post-quench Hamiltonian $H_i$. Now we are in a position to write $\langle \mathcal{O}^x_{lm}(t)\rangle_0$ as the ground state expectation value of a product of an even number of fermionic operators and then by applying Wick's theorem, we can reduce it to a sum of elementary contraction involving ground state expectation values of different pairs of operators~\cite{Uhrich2020}. For this we need to calculate the expectation values of four types of pairs: $\langle B^-_l A^+_m \rangle_0$, $\langle A^+_l B^-_m \rangle_0$, $\langle B^-_l B^-_m \rangle_0$, and $\langle A^+_l A^+_m \rangle_0$, see appendix \ref{app:ssh} for details.

Out of these four types of expectation values, two turn out to be complex. Therefore we calculate the absolute value of the determinant of the following $2(m-l)$ dimensional complex matrix and investigate its dynamics:
\begin{equation}
\mathcal{O}^x_{l,m}(t)=
\begin{vmatrix}
F_{l,l+1}(t) & F_{l,l+2}(t) & \cdots & F_{l,m}(t) \\
F_{l+1,l+1}(t) & \cdots & \cdots & F_{l+1,m}(t) \\
\vdots & & & \vdots \\
F_{m-1,l+1}(t) & \cdots & \cdots & F_{m-1,m}(t)
\end{vmatrix}
\end{equation}
where
\begin{equation}
F_{l,m}(t)=\begin{pmatrix}
   \langle B^-_{l}(t)B^-_{m-1}(t)\rangle_0 &  \langle B^-_{l}(t)A^+_{m}(t)  \rangle_0\\
     \langle A^+_{l+1}(t)B^-_{m-1}(t)\rangle_0 &  \langle A^+_{l+1}(t)A^+_{m}(t)\rangle_0
\end{pmatrix}\,.
\end{equation}
We note that $\langle B^-_i B^-_i \rangle_0 = \langle A^+_i A^+_i \rangle_0 = 0$ and that the matrix $\mathcal{O}^x_{l,m}(t)$ is anti-symmetric. In the general non-equilibrium case the string order parameter is the Pfaffian of the matrix $\mathcal{O}^x_{l,m}(t)$. However as the matrix is anti-symmetric we can calculate the square-root of the determinant~\cite{Uhrich2020}. As in the equilibrium case we define
\begin{equation}
\mathcal{M}(t)=\sqrt{\left|\mathcal{O}^x_{1,\frac{N}{2}}(t)\right|}\,.
\end{equation}
We want to point out that following from these definitions $\mathcal{M}(t=0)=\mathcal{M}$, however $\mathcal{O}^x_{l,m}(t=0)$ is not identical to the equilibrium case, which is defined in its simplest possible form. The results for the dynamics of the string order parameter are discussed in section \ref{sec:results}.

\section{String Order Parameter for an extended Kitaev chain}\label{lrkc}

\subsection{The Model}\label{sec:kitmodel}

To check more generally the dynamics of the string order parameter and check relations between different possible measures of the topology, we consider also an extended Kitaev chain~\cite{Kitaev2001,Sedlmayr2018} with several extra longer-range hopping terms:
\begin{align}\label{khlr}
H=\sum_{i>j}&\Psi^\dagger_{i}\left(
\Delta_{|i-j|}i{\bm\tau}^y-J_{|i-j|}{\bm\tau}^z\right)\Psi_{j}+\textrm{H.c.}\\\nonumber&-\mu\sum_{j}\Psi^\dagger_{j}{\bm\tau}^z\Psi_{j}\,.
\end{align}
The operators are $\Psi^\dagger_{j}=(c^\dagger_{j},c_{j})$, where $c_{j}^{(\dagger)}$ annihilates (creates) a spinless fermionic particle at a site $j$. The long-range hopping is $\vec J$ and p-wave hopping $\vec\Delta$ which can both be truncated at a suitable distance $R$. $J_1=J$ is the nearest neighbor hopping which is used as the overall energy scale, $\mu$ is the chemical potential and $\vec{\tau}=(\tau^x,\tau^y,\tau^z)^T$ are Pauli spin matrices. The benefit of this model is that it allows for the tuning of the winding number $\nu$ to in principle any desired value (see appendix \ref{app:top}), but it does not require truly long range coupling, reducing problems from finite size effects. This model remains still a relatively simple two-band topological model in class BDI~\cite{Kitaev2009,Ryu2010,Teo2010}, and therefore gives us a good balance between simplicity and complexity for analyzing the dynamics. We will focus on the string order parameters for the $\nu=\pm1$ phases, though we will quench between a wider variety of initial states and Hamiltonians.

After a Fourier transform the Hamiltonian becomes
\begin{equation}
H = \frac{1}{2}\sum\limits_{k} \Psi_k^{\dagger} \mathcal{H}_k \Psi_k
\end{equation}
with $\Psi_k^\dagger \equiv (c_k^\dagger, c_{-k})$. The Hamiltonian density is
\begin{equation}
\mathcal{H}_k \equiv\vec{d}_k \cdot\vec{\tau}
\label{ekc-hk}
\end{equation}
where
\begin{equation}
\vec{d}_k = \begin{pmatrix}
    0\\
    2\sum_{m=1}^{R} \Delta_m \sin mk\\
    -2\sum_{m=1}^{R} J_m \cos mk - \mu 
\end{pmatrix}\,.
\end{equation}
The eigenvalues of $\mathcal{H}_k$ are $\mp \varepsilon_k = \mp|d_k|$. More details on the model and its diagonalization can be found in appendix \ref{app:kit}.

\subsection{The String Order Parameter}\label{sec:kitsop}

\begin{figure}
\includegraphics[width=0.95\columnwidth]{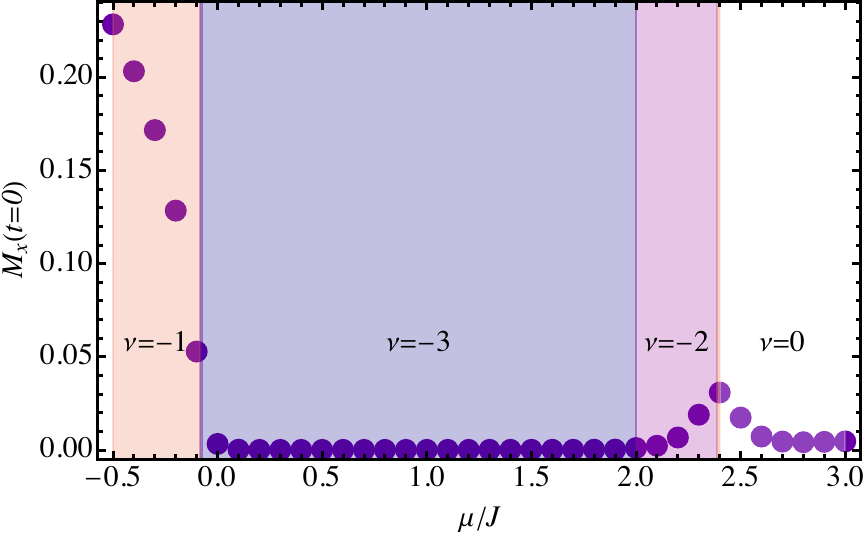}
\caption{Plots of $\mathcal{M}_x$ vs $\mu$ for $N=1000$ for the ground state of the extended Kitaev model with $\vec{J}=(1,1,1)J$, $\vec\Delta=(1.2,1.4,1.6)J$. In the $\nu=-1$ phase the string order parameter $\mathcal{M}_x\neq0$ as expected. In all other phases it is zero. We note though that near the gap-closing point between $\nu=-2$ and $\nu=0$ finite size effects are large and $\mathcal{M}_x$ has still not fully converged to zero. In Fig.~\ref{lrkc-N} we demonstrate the scaling of $\mathcal{M}_x$ for the representative points in the phase diagram we will use in the quenches.}
\label{fig:kiteq}
\end{figure}

The extended Kitaev model has similar string order parameters to the SSH model for the phases $\nu=\pm1$~\cite{Uhrich2020}. For $\nu=-1$ we can define
\begin{equation}
\mathcal{O}^x_{lm} = \left\langle  \Phi^{-}_{l} \left[ \prod\limits_{j=l+1}^{m-1} \Phi^{+}_{j} \Phi^{-}_{j} \right] \Phi^{+}_{m}\right\rangle_0
\end{equation}
where $\Phi^{\pm}_{j} = c^{\dagger}_j \pm c_j$. For $\nu=1$ the appropriate string order parameter is
\begin{equation}
\mathcal{O}^y_{lm} = \left\langle \Phi^{+}_{l} \left[ \prod\limits_{j=l+1}^{m-1} \Phi^{-}_{j} \Phi^{+}_{j} \right] \Phi^{-}_{m} \right\rangle_0 \,.
\end{equation}
We again consider a quench scenario, with the average over the initial state $|\Psi_i \rangle$, the half-filled ground state of the Hamiltonian $H_i$. From time $t=0$ we time-evolve with a Hamiltonian $H_f$. We focus on two sets of quenches. The first is for $R=1$, \emph{i.e.}~the usual Kitaev chain, with $\Delta=\pm 1.2J$ and $\mu\in\{1.75,2.25\}J$. If $|\mu|\leq2J$ then $\nu=-\textrm{sgn}(\Delta)$, otherwise it is zero. The second uses $R=3$, $\vec{J}=(1,1,1)J$, and $\vec\Delta=(1.2,1.4,1.6)J$, see Fig.~\ref{fig:kiteq}. The chemical potential is varied between $\mu\in\{3,-0.5,2.2,0.9\}J$ for winding numbers $\nu\in\{0,-1,-2,-3\}$ respectively.

At time $t\geq0$ we find
\begin{equation}
\mathcal{O}^x_{lm}(t) = \left\langle  \Phi^{-}_{l}(t) \left[ \prod\limits_{j=l+1}^{m-1} \Phi^{+}_{j}(t) \Phi^{-}_{j}(t) \right] \Phi^{+}_{m}(t)\right\rangle_0 \,.
\end{equation}
After applying Wick's theorem this results in
\begin{equation}\label{det}
\mathcal{O}^{x,y}_{lm}(t) =
\begin{vmatrix}
F^{x,y}_{l,l+1}(t) & F^{x,y}_{l,l+2}(t) & \cdots & F^{x,y}_{l,m}(t) \\
F^{x,y}_{l+1,l+1}(t) & \cdots & \cdots & F^{x,y}_{l+1,m}(t) \\
\vdots & & & \vdots \\
F^{x,y}_{m-1,l+1}(t) & \cdots & \cdots & F^{x,y}_{m-1,m}(t)
\end{vmatrix}
\end{equation}
where
\begin{equation}\label{kitfm}
F^x_{l,m}(t)=
    \begin{pmatrix}
        \langle \Phi^{-}_{l}(t)\Phi^{-}_{m-1}(t)\rangle_0 & \langle \Phi^{-}_{l}(t)\Phi^{+}_{m}(t)\rangle_0\\
        \langle \Phi^{+}_{l+1}(t)\Phi^{-}_{m-1}(t)\rangle_0 & \langle \Phi^{+}_{l+1}(t)\Phi^{+}_{m}(t)\rangle_0
    \end{pmatrix}\,.
\end{equation}
and for $O^{y}_{lm}$, the determinant will follow from Eq.~(\ref{det}) with the definition of $F_{l,m}(t)$ altered as
\begin{equation}\label{kitfm_y}
F^y_{l,m}(t)=
    \begin{pmatrix}
        \langle \Phi^{+}_{l}(t)\Phi^{+}_{m-1}(t)\rangle_0 & \langle \Phi^{+}_{l}(t)\Phi^{-}_{m}(t)\rangle_0\\
        \langle \Phi^{-}_{l+1}(t)\Phi^{+}_{m-1}(t)\rangle_0 & \langle \Phi^{-}_{l+1}(t)\Phi^{-}_{m}(t)\rangle_0
    \end{pmatrix}\,.
\end{equation}
The results for the correlation functions are given in appendix \ref{app:kit}.

\begin{figure}
\includegraphics[width=0.95\columnwidth]{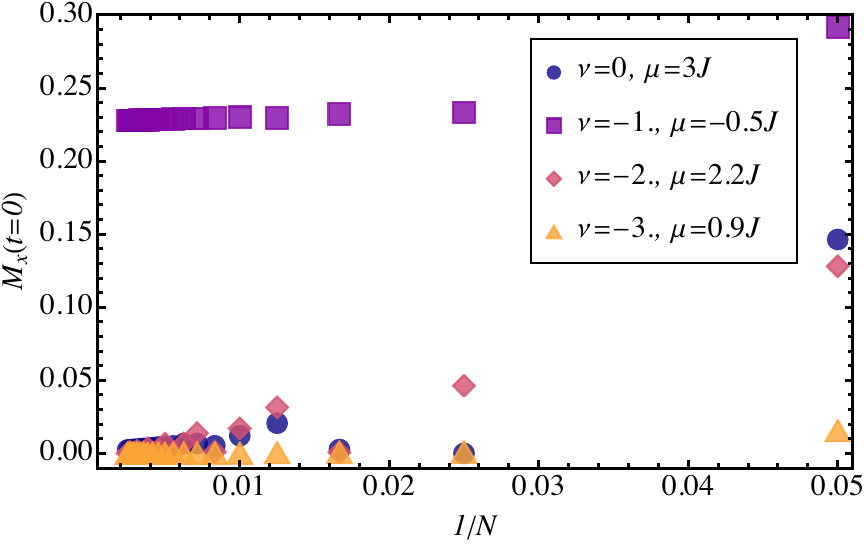}
\caption{Finite size scaling of the string order parameter $\mathcal{M}_x$ in the ground state for $\vec{J}=(1,1,1)J$ and $\vec\Delta=(1.2,1.4,1.6)J$. The chemical potential is changed to reach examples of various phases, as marked on the figure. As expected $\mathcal{M}_x$ scales to zero except in the $\nu=-1$ topological phase. Typically everything has converged by $1/N\sim0.005$, \emph{i.e.}~$N\sim200$.}
\label{lrkc-N}
\end{figure}

As before we define
\begin{equation}
\mathcal{M}_{x,y}(t)=\sqrt{\left|\mathcal{O}^{x,y}_{1,\frac{N}{2}}(t)\right|}\,.
\end{equation}
To demonstrate the behavior of the string order parameter in equilibrium we take a cut through a part of the very large parameter space for $R=3$. In Fig.~\ref{fig:kiteq} we plot the string order parameter as a function of chemical potential across phases $\nu\in\{-1,-3,-2,0\}$. It is only non-zero in the $\nu=-1$ phase. For $\mathcal{M}^y$ we would find a similar result with non-zero values in the $\nu=1$ phase. The dependence of the string order parameter on system-size is demonstrated in Fig.~\ref{lrkc-N}.

\section{Critical Times, Entanglement Entropy, and Entanglement Spectra}\label{sec:ee}

Our aim is to establish how general a connection can be drawn between different concepts of dynamical criticality or phases. Therefore we wish to compare the results for the dynamical string order parameter to the dynamical entanglement entropy and spectra, and the critical times associated with dynamical quantum phase transitions. Dynamical quantum phase transitions are defined as non-analyticities in the return rate, caused by zeroes in the Loschmidt amplitude which occur as a function of time~\cite{Heyl2013}. The Loschmidt amplitude is given by
\begin{equation}
    L(t)=\langle\Psi_i|e^{iH_ft}|\Psi_i\rangle
\end{equation}
and the return rate in the thermodynamic limit is
\begin{equation}
    l(t)=-\lim_{N\to\infty}\frac{1}{N}\ln|L(t)|\,.
\end{equation}
The zeroes of the Loschmidt amplitude can be understood as Fisher zeroes in the complex time plane~\cite{Heyl2013,Maslowski2024b}. When the zeroes cross the real time axis there is a critical time corresponding to the non-analyticity. For the models we are interested in the critical times can be easily calculated as~\cite{Vajna2014}:
\begin{equation}
t_{m,n}(k_c) = \frac{\pi(2n + 1)}{2\varepsilon^f_{k_{cm}}}
\label{tc-def}
\end{equation}
where $\varepsilon^f_{(k_{cm})}$ is the eigenenergy for $H_f$ evaluated at the critical momentum $k_{cm}$ defined by
\begin{equation}
    \vec{d}_{k_{cm}}^i\cdot\vec{d}^f_{k_{cm}}=0\,.
\end{equation}
The vectors $\vec{d}^{i,f}_{k}$ parameterize the pre- and post-quench Hamiltonians, see Eq.~\eqref{ekc-hk}. More general cases can be found in Refs.~\onlinecite{Vajna2014,Sedlmayr2018,Maslowski2020,Maslowski2024b,Yang2026}. DQPTs for these models have been well studied, and here we take only the critical times. Note that there may be multiple critical times both due to their periodicity, $n$, and multiple critical momenta being possible, labeled by $m$.

The ground states of both considered models can be completely characterized by their respective correlation matrices, which in our case are time dependent. Correlation matrices carry indices for position in the chain ($l,m$) and the local basis given explicitly in the following,
\begin{equation}\label{correl_mats}
\begin{aligned}
C^{SSH}_{lm}(t)&=
    \begin{pmatrix}
        \big\langle a^\dagger_l(t) a_m(t)\big\rangle_0 & \big\langle a^\dagger_l(t) b_m(t) \big\rangle_0\\
        \big\langle b^\dagger_l (t)a_m(t)\big\rangle_0 & \big\langle b^\dagger_l(t) b_m(t)\big\rangle_0
    \end{pmatrix}\,,\textrm{ and} \\
C^{EK}_{lm}(t)&=
    \begin{pmatrix}
        \big\langle c^\dagger_l(t) c_m(t)\big\rangle_0 & \big\langle c^\dagger_l(t) c^\dagger_m(t) \big\rangle_0\\
        \big\langle c_l(t)c_m(t)\big\rangle_0 & \big\langle c_l(t) c^\dagger_m(t)\big\rangle_0
    \end{pmatrix}\,,
\end{aligned}
\end{equation}
for the SSH and extended Kitaev models respectively. Limiting $l,m$ to a subsystem $\Omega$, such  a sub-block of the correlation matrix has eigenvalues $\lambda_n(t)$ which can be used to find the entanglement entropy and constitute the single particle entanglement spectrum. Crossings at $\lambda_n(t^*)=0.5$ are usually associated with critical times~\cite{Gong2018a,Poyhonen2021}. Defining the entanglement Hamiltonian $\mathcal{H}_E$ via the time dependent density matrix $\rho(t)=e^{-\mathcal{H}_E}(t)/\mathcal{Z}$, the eigenvalues of the entanglement Hamiltonian $\xi_n$ are related to the entanglement spectrum by $\xi_n(t)=\ln[(1-\lambda_n(t))/\lambda_n(t)]$. Therefore $\lambda_n(t^*)=0.5$ leads to a crossing at $\xi_n(t^*)=0$~\cite{Peschel2003}.

The entanglement spectrum is then used to calculate the von Neumann entanglement entropy of the subsystem $\Omega$ and its complement $\bar{\Omega}$~\cite{Bennett1996,Vidal2003a, Sirker2014}. Here we always choose one half of the chain as the subsystem \emph{i.e.}~$|\Omega| = N/2$, with $N=200$ unit cells for the SSH model, and $N=350$ sites for the Kitaev chain. The entanglement entropy is
\begin{equation}\label{vonNeumann}
S(t) = -\sum_n \left[ \lambda_n(t) \ln\left[\lambda_n(t)\right] + \left[1-\lambda_n(t)\right]\ln\left[1-\lambda_n(t)\right] \right]\,.
\end{equation}
In this article we will not explicitly consider the entanglement entropy. We note that constructing the time-dependent correlation matrix in a quench scenario is a standard task~\cite{Poyhonen2021,Micallo2020}.

\section{Results for Dynamics in the SSH Model}\label{sec:results}

\begin{figure}
\includegraphics[width=0.95\columnwidth]{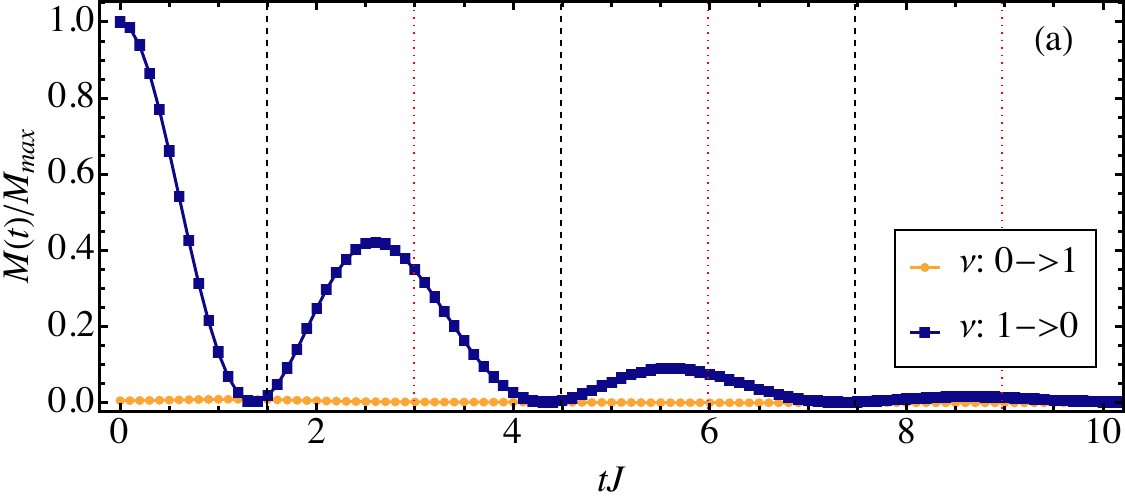}\\
\includegraphics[width=0.95\columnwidth]{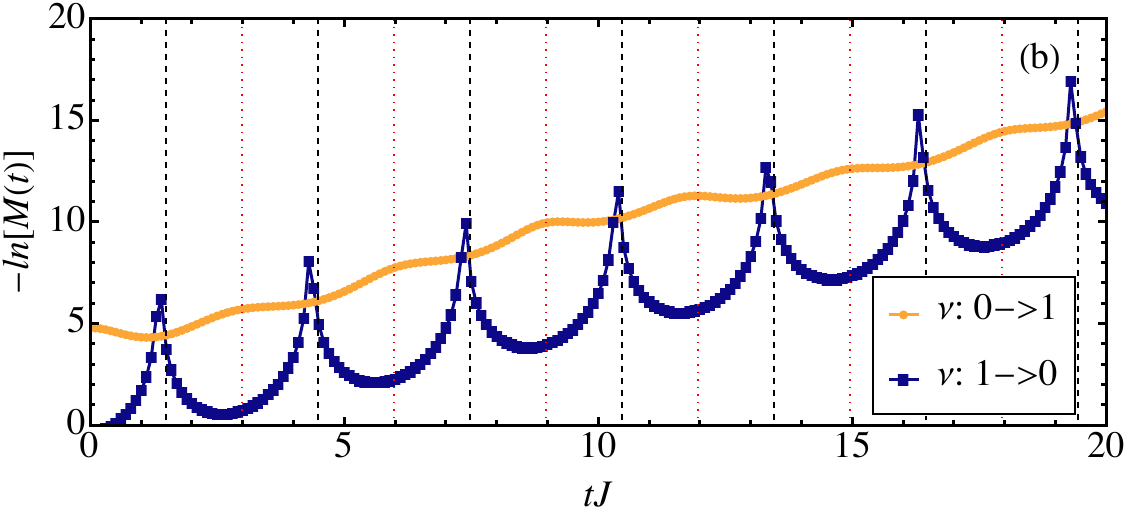}
\caption{Plots of $\mathcal{M}(t)$ for quenches in the SSH model with $|\delta|=0.4$, as marked on the panels. The system size is $N=200$. Vertical dashed black lines show the DQPT critical times and red dotted lines are plotted half way between the critical times as a visual aid. Panel (a) is the string order parameter rescaled by its maximum value, and panel (b) is the $\ln$ of the string order parameter. Zeroes are clearly visible in the non-trivial to trivial quench, $\nu:1\to0$ where $\delta:0.4\to-0.4$. The zeroes occur approximately at the DQPT critical times. The opposite quench shows small oscillations with a similar period, but no zeroes.}
\label{sopx-neq-t}
\end{figure}

Before considering some of the more complicated quenches possible in the extended Kitaev chain, we consider the exemplary SSH model. We note that $\mathcal{M}(t)$ has already converged for $N\gtrsim 200$. In Fig.~\ref{sopx-neq-t} we plot $\mathcal{M}(t)$ for two example quenches between the topologically trivial and non-trivial phases with $N=200$. For the quench from the non-trivial phase to the trivial phase, $\delta: 0.4\to-0.4$, $\mathcal{M}(t=0)\neq0$ and passes through zero periodically, close to the DQPT critical time, decaying eventually to zero in the long time limit. For the opposite case, $\delta: -0.4\to0.4$, we naturally have $\mathcal{M}(t=0)=0$ suggesting perhaps that there are then no interesting dynamics of this quantity for such a quench. However, by considering $\ln[\mathcal{M}(t)]$ we can observe some small oscillations with a similar period as for the opposite quench.

\begin{figure}
\includegraphics[width=0.95\columnwidth]{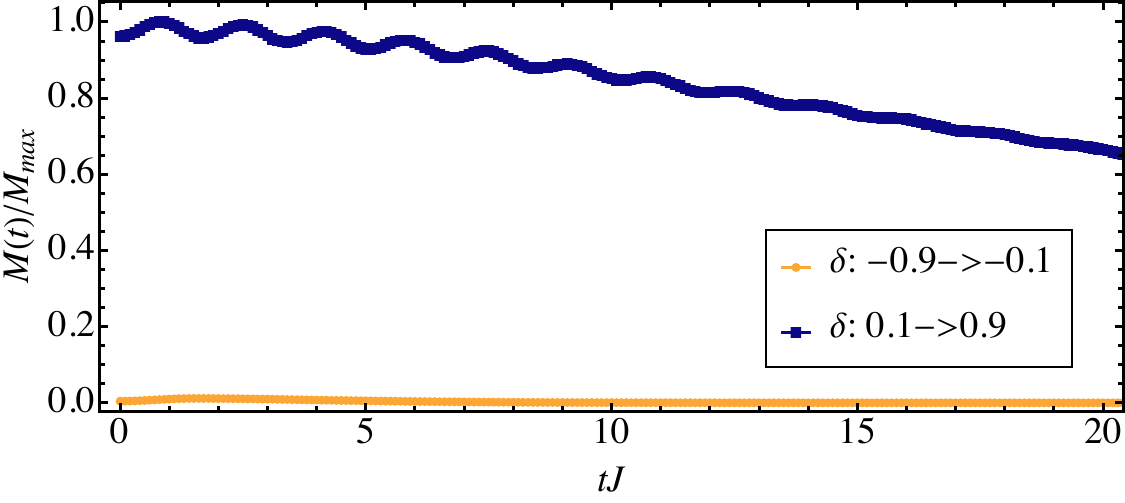}
\caption{Plots of $\mathcal{M}(t)$ for quenches in the SSH model within topological phases, as marked on the panels. The system size is $N=200$. Small oscillations can be seen, but no zeroes occur and similarly there are no DQPTs, and hence no such critical times.}
\label{sopx-neq-t-2}
\end{figure}

\begin{figure}
\includegraphics[width=0.95\columnwidth]{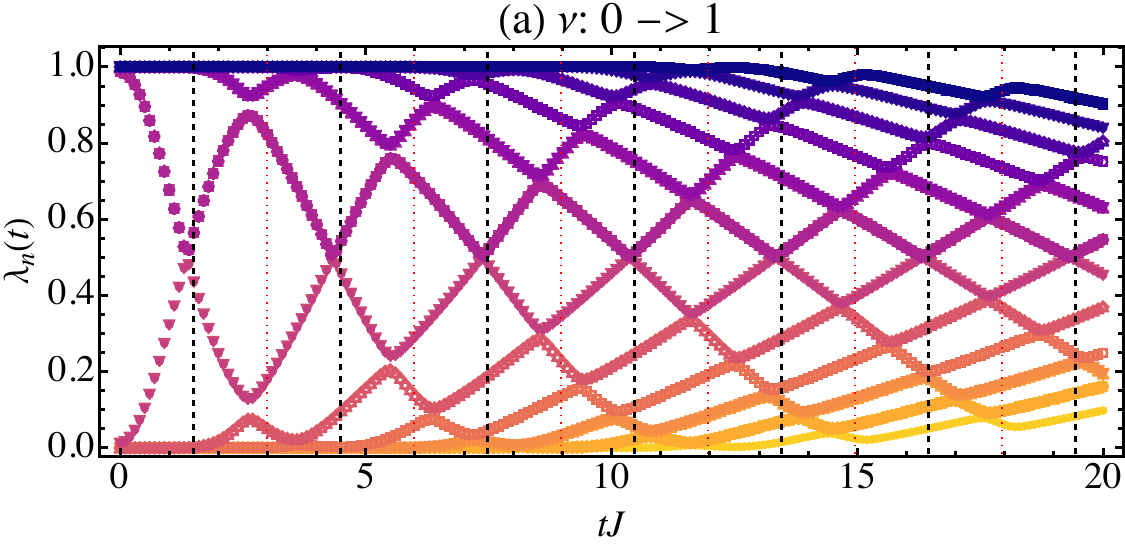}\\
\includegraphics[width=0.95\columnwidth]{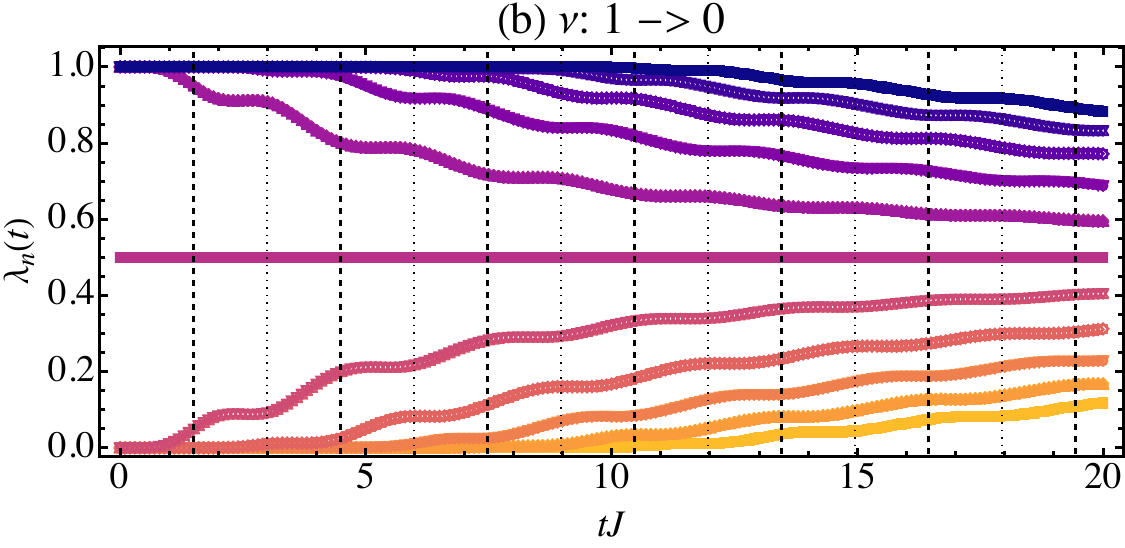}
\caption{Plots of the entanglement spectra for quenches in the SSH model with $|\delta|=0.4$, as marked on the panels. The 22 eigenvalues $\lambda_n(t)$ closest to $0.5$ are shown. The system size is $N=200$. Vertical dashed black lines show the DQPT critical times and red dotted lines are plotted half way between the critical times as a visual aid. For the quench $\nu:0\to1$ in panel (a) crossings of $\lambda_n(t)$ at $0.5$ can be seen close to the DQPT critical times, see Fig.~\ref{fig:app-ssh-es-sc} for a close up. For the opposite quench there are oscillations but no crossings at $0.5$.}
\label{fig:ssh-es}
\end{figure}

Small oscillations in $\mathcal{M}(t)$ can also be seen if we quench within the topologically non-trivial phase, see Fig.~\ref{sopx-neq-t-2}. In this case there are no DQPTs and no zeroes of the string order parameter, as may be expected. If we compare now to the dynamical behavior of the entanglement spectra, see Fig.~\ref{fig:ssh-es}, we can observe a different scenario. In this case we see crossings of the entanglement spectrum at $0.5$ for the quench from the topologically trivial phase to the non-trivial phase. The crossings occur close to the DQPT critical times. For the opposite quench there are oscillations with a similar period, but no crossings at $0.5$.

From these results we can note two things. When critical phenomena occur, by which we mean; DQPT critical times, $\lambda_n(t)$ crossings at $0.5$, and string order parameter zeroes of $\mathcal{M}(t)$; they occur at very similar times. More generally any oscillatory behavior has a similar time scale. As in this case the bulk DQPT behavior is identical for the quenches $\nu:0\to1$ and $\nu:1\to0$ there can not be a direct connection to the different behavior observed for the string order parameter and entanglement seen here.

To check more carefully if the critical times coincide we can check the scaling. Focusing on the first critical time for the quenches $\nu:0\to1$ and $\nu:1\to0$ at $|\delta|=0.4$ we can find that the DQPT critical time is in both cases $t_c=1.495/J$. The entanglement spectra have a crossing $\lambda_n(t*)=0.5$ at $t^*=1.36/J$ (for $\nu:0\to1$). See appendix \ref{app:scaling} for more details. The string order parameter critical time $\mathcal{M}(t')=0$ (for $\nu:1\to0$ in this case) coincidentally occurs also at $t'=1.36/J$. This is a result due to the simplicity of the SSH model and does not generalize further, we discuss it briefly in appendix \ref{app:scaling}. In the next section we will check more widely how the critical times differ for more general quenches.

It is perhaps not surprising that these different types of criticality do not in fact perfectly coincide. The origin of each is distinct, and purely from a mathematical point of view it is hard to see how a zero of the string order parameter should occur at anything like the same time as the zero of the Loschmidt echo. Nonetheless what is seen here, and in other studies, is that these critical times are close together. This itself demands an explanation.

Oscillations in entanglement entropy can be understood as caused by restructuring of the correlations between sites following the quench~\cite{Sedlmayr2018}. For example one of the entanglement eigenvalues $\lambda_n(t)=0.5$ in Fig.~\ref{fig:ssh-es}(b) corresponds to the entanglement between the edge modes of the initial state. For Fig.~\ref{fig:ssh-es}(a) the initial state has no such entanglement as there is no edge mode at $t=0$, however during the dynamics the correlations are restructured until a perfect dimer is formed at a critical time giving $\lambda_n(t)=0.5$. How is this related to the DQPT $t_c$? The formation of such dimers for the SSH model also represents the point at which the overlap between such a state and the initial one can vanish. Clearly however these two do not necessarily have to happen close in time, though this argument gives a physical reason why they may.

For the string order parameter zeroes we must start in the topologically non-trivial phase. A zero thus represents the time at which the string order has completely vanished and the system appears completely trivial in this sense. Again this may typically happen when the correlations have been completely restructured, a time scale closely related to, but not identical to, that of the DQPT critical time. In the next section we consider a more complex model which can show the limitations of these connections.

\section{Results for Dynamics in the Extended Kitaev Chain}\label{sec:kitres}

In this section we consider quenches in the extended Kitaev chain. The first cases are for $R=1$, the normal nearest neighbor Kitaev chain. In this case we can have phases $\nu\in\{-1,0,1\}$ and we have two distinct string order parameters for the two topologically non-trivial phases, $\mathcal{M}_{x,y}$ for $\nu\in\{-1,1\}$ respectively. We then add longer range hopping and pairing terms, allowing us to tune to a wider variety of topological phases, but we remain focused on the same string order parameters. In principle many different quenches between different phases are now possible, and we will focus on here on some exemplary cases. For example swapping  $\nu:1\longleftrightarrow-1$ and $\mathcal{M}_{y}\longleftrightarrow\mathcal{M}_{x}$ leaves results the same, all else being equal.

\begin{figure}
\includegraphics[width=0.95\columnwidth]{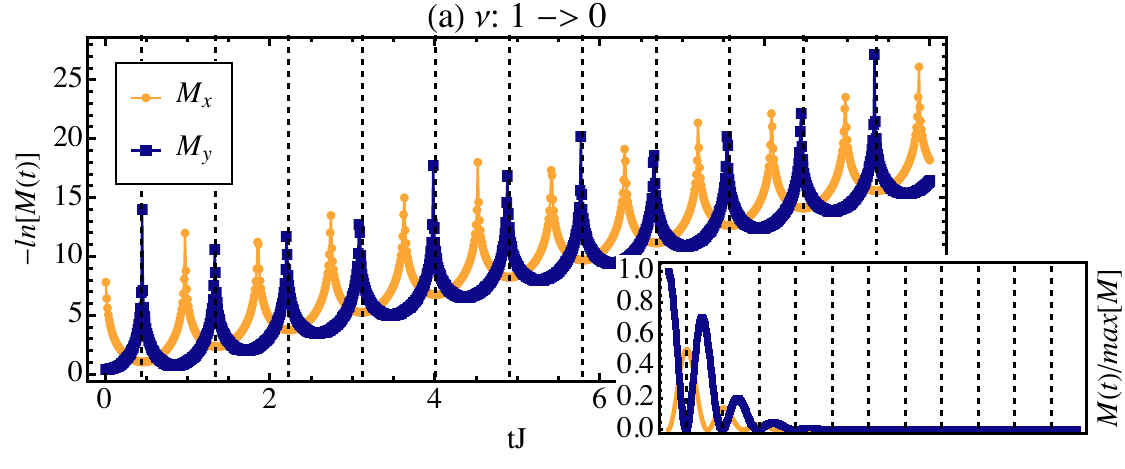}\\
\includegraphics[width=0.95\columnwidth]{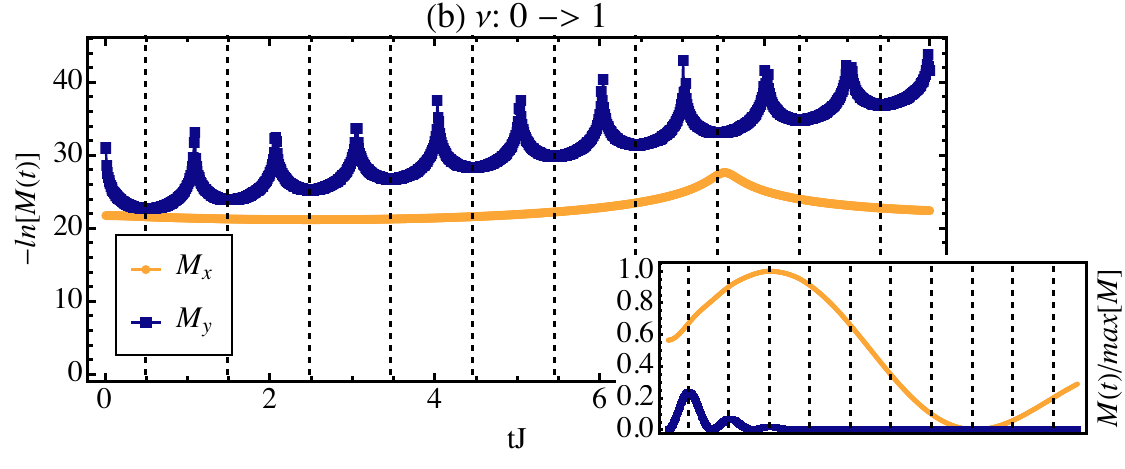}\\
\includegraphics[width=0.95\columnwidth]{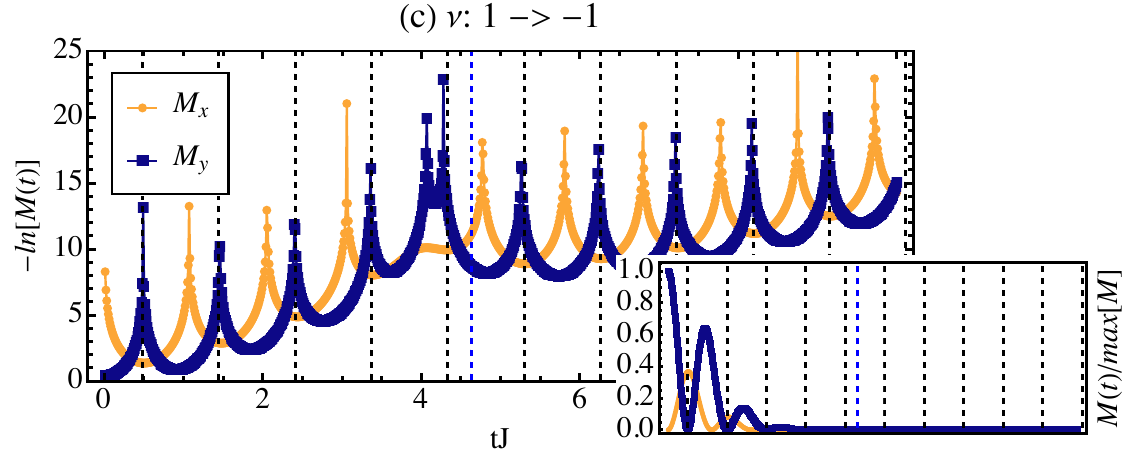}
\caption{Plots of $\mathcal{M}_{x,y}(t)$ for different quenches, as labeled on the figures. Results are for the Kitaev chain with $R=1$ and $N=350$, see Sec.~\ref{sec:kitmodel} for details of the parameters. Insets show $\mathcal{M}_{x,y}(t)$ rescaled by its maximum and main panels show the log to highlight location of the zeroes. Interesting to note in panel (a) is that both $\mathcal{M}_x(t)$ and $\mathcal{M}_y(t)$ have some interesting dynamics, despite the quench being $\nu:1\to0$, \emph{i.e.}~for two phases where $\mathcal{M}_x=0$ in both. See main text for further discussion. Vertical dashed lines show the DQPT critical times, with different colors referring to distinct critical momenta.}
\label{lrkc-t3}
\end{figure}

Fig.~\ref{lrkc-t3} shows both $\mathcal{M}_x(t)$ and $\mathcal{M}_y(t)$ for several simple quenches in the $R=1$ Kitaev chain. For $\nu:1\to0$ $\mathcal{M}_y(t)$ starts at a non-zero value and decays, with zeroes close to the DQPT critical times. This is directly analogous to what is seen for the SSH model in the preceding section. Interestingly $\mathcal{M}_x(t)$, which is zero for both the $\nu=0$ and $\nu=1$ topological phases, also displays some interesting behavior. Starting at $\mathcal{M}_x(0)=0$ it becomes non-zero and then also decays with zeroes forming in between the $\mathcal{M}_y(t)$ zeroes. This pattern repeats for the time scale we can follow, though there seems to be no exact relation between the location of the zeroes. Quenching in the opposite direction $\nu:0\to1$ we find that $\mathcal{M}_y(t)$ has critical times in between the DQPT critical times. Such behavior was not observed in the SSH model. In this case $\mathcal{M}_x(t)$ shows no clear behavior as may be expected. Finally we can consider a quench $\nu:1\to-1$, \emph{i.e.}~between the $\mathcal{M}_y$ and $\mathcal{M}_x$ phases. for the DQPT there are now two distinct sets of critical times. Close to the second DQPT critical time an additional zero can be observed in $\mathcal{M}_y(t)$, though again there is no exact relation between their numerical values. We also note the similar qualitative behavior for both $\nu:1\to0$ in Fig.~\ref{lrkc-t3}(a) and $\nu:1\to-1$ in Fig.~\ref{lrkc-t3}(c). Even taking both $\mathcal{M}_x(t)$ and $\mathcal{M}_y(t)$ into account does not distinguish clearly these two types of quench. One general behavior which can be observed is the appearance of double zeroes in the relevant string order parameter when there are nearby different critical times of the DQPT, see Fig.~\ref{lrkc-t3}(c).

\begin{figure}
\includegraphics[width=0.95\columnwidth]{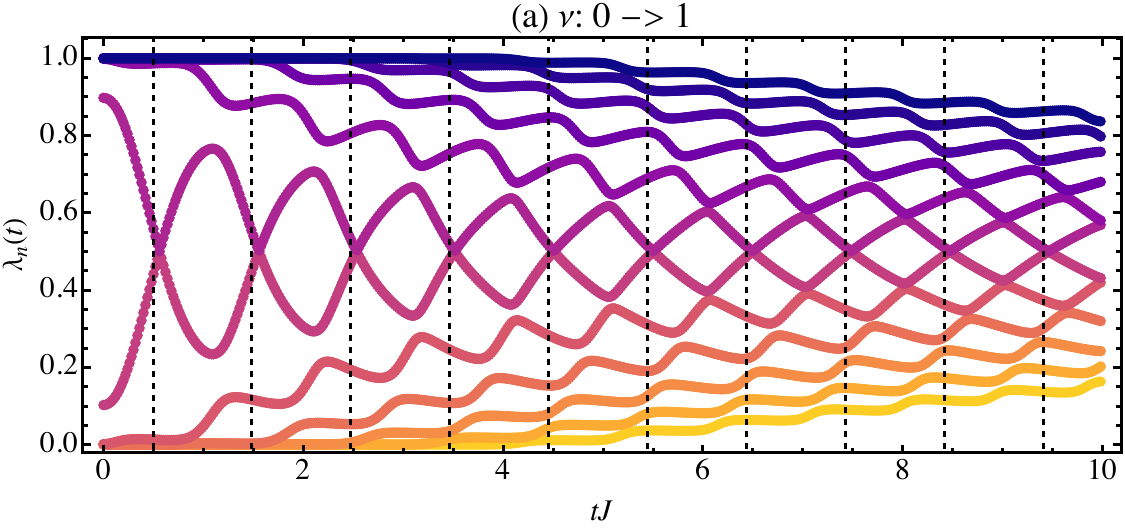}\\
\includegraphics[width=0.95\columnwidth]{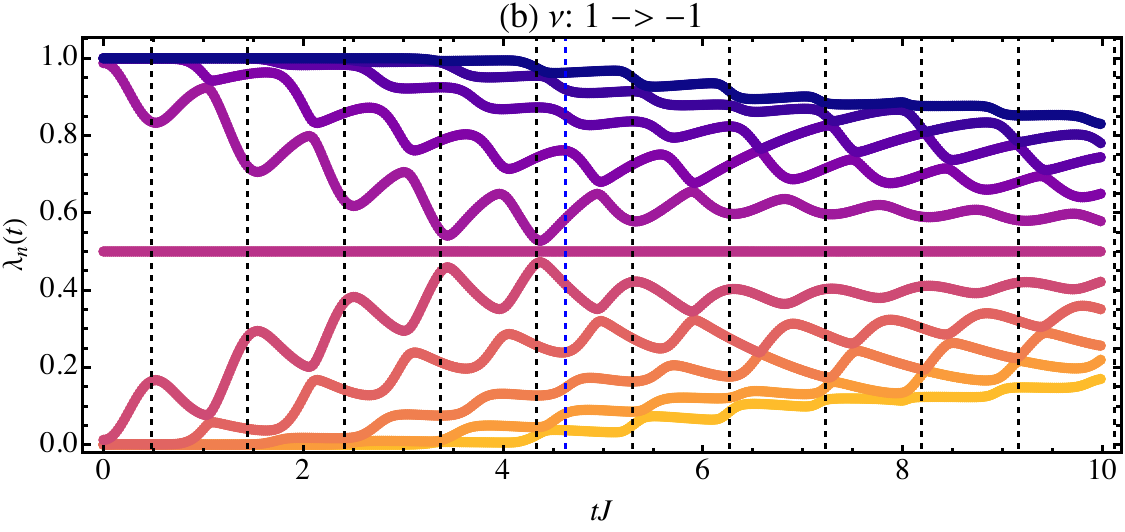}\\
\includegraphics[width=0.95\columnwidth]{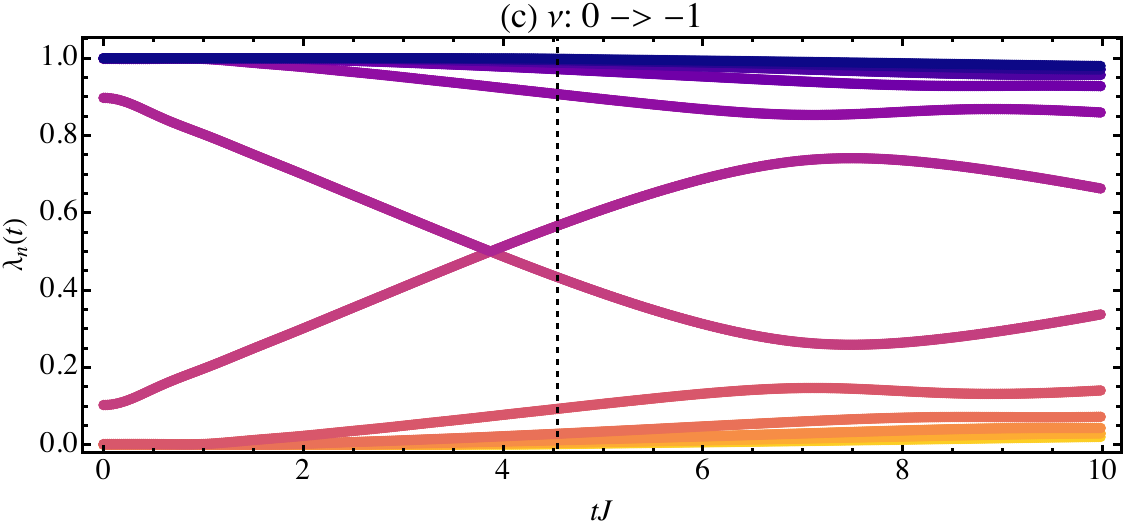}
\caption{Plots of the entanglement spectra for different quenches, as labeled on the figures. The 22 eigenvalues $\lambda_n(t)$ closest to $0.5$ are shown. Results are for the Kitaev chain with $R=1$ and $N=350$, see Sec.~\ref{sec:kitmodel} for details of the parameters. Vertical dashed lines show the DQPT critical times, with different colors referring to distinct critical momenta.}
\label{lrkc-es-t3}
\end{figure}

The behavior of the string order parameter can be compared to that of the entanglement spectrum. For $\nu:1\to0$ we see only oscillations, with no crossings at $0.5$ (not shown here). In Fig.~\ref{lrkc-es-t3} we show $\lambda_n(t)$ for some of the same quenches as in \ref{lrkc-t3}. For quenches into topologically non-trivial phases crossings can be seen. For the two simple quenches $\nu:0\to\pm1$ they are close to the DQPT critical times. For the quench $\nu:1\to-1$ we note strong oscillations on the time scale of one critical time, and crossings on the time scale of the second critical time of the DQPTs, see Fig.~\ref{lrkc-es-t3}(b). In this case there is also a pair of $\lambda_n(t)$ pinned to $0.5$ from the initial state entanglement spectrum.

\begin{figure}
\includegraphics[width=0.95\columnwidth]{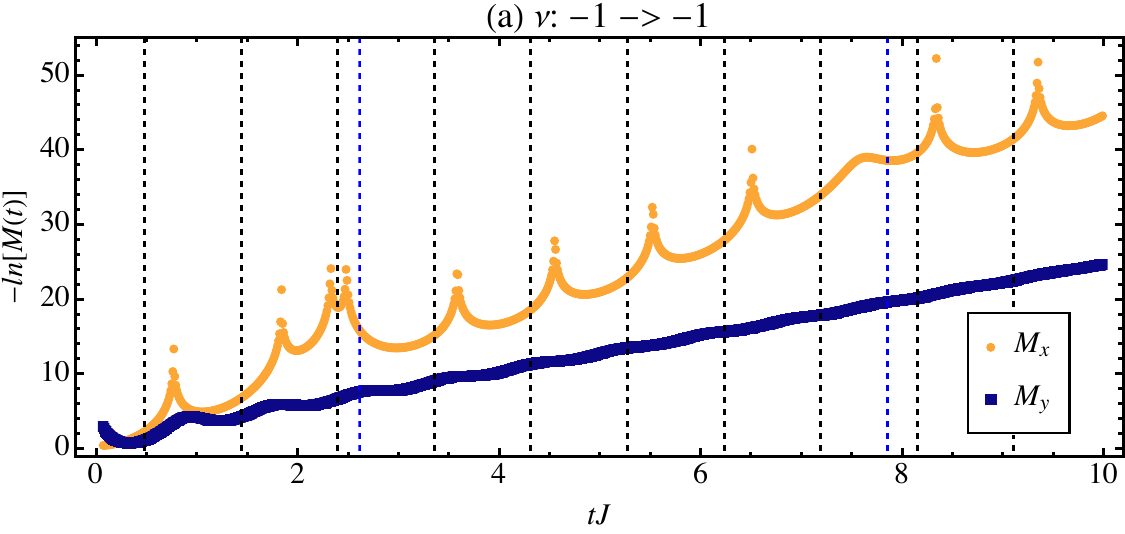}\\
\includegraphics[width=0.95\columnwidth]{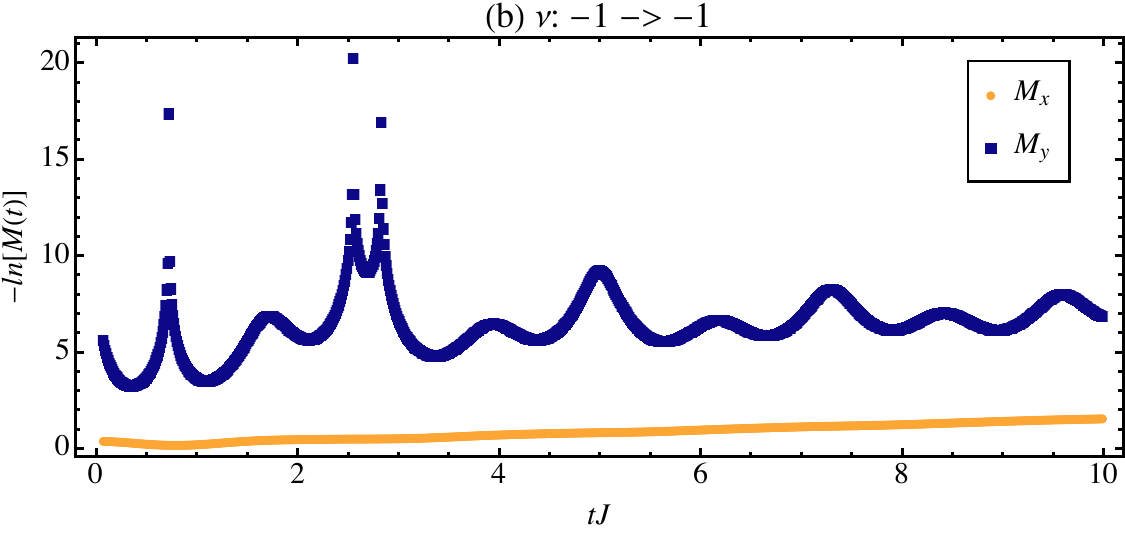}
\caption{Plots of $\mathcal{M}_{x,y}(t)$ for different types of quenches within the phase $\nu=-1$ of the Kitaev chain with $R=1$ and $N=350$. In panel (a) we quench from $(\Delta,\mu)=(1.2,1.75)J$ to $(\Delta,\mu)=(0.4,-1.75)J$ and in panel (b) we quench from $(\Delta,\mu)=(1.2,1.75)J$ to $(\Delta,\mu)=(1.2,0.75)J$. DQPT critical times, when present, are shown as vertical dashed lines, as for Fig.~\ref{lrkc-t3}.}
\label{lrkc-t4}
\end{figure}

An additional case we can consider is quenching inside a phase. For the extended Kitaev chain we can find cases with and without DQPTs for such quenches, see Fig.~\ref{lrkc-t4}. In both cases shown we quench within the $\nu=-1$ phase where $\mathcal{M}_x\neq0$. When DQPTs are present zeroes can be seen in $\mathcal{M}_x(t)$, when no DQPTs are present there are in any case zeroes in $\mathcal{M}_y(t)$, though they clearly do not form any consistent structure. In general what we see is that although dynamical string order parameter zeroes are often near the critical times of DQPTs, there is in general no direct connection between the quenched phases, the relevant string order parameter which displays zeroes, or the exact times of the critical points.

\begin{figure*}
\includegraphics[width=0.95\columnwidth]{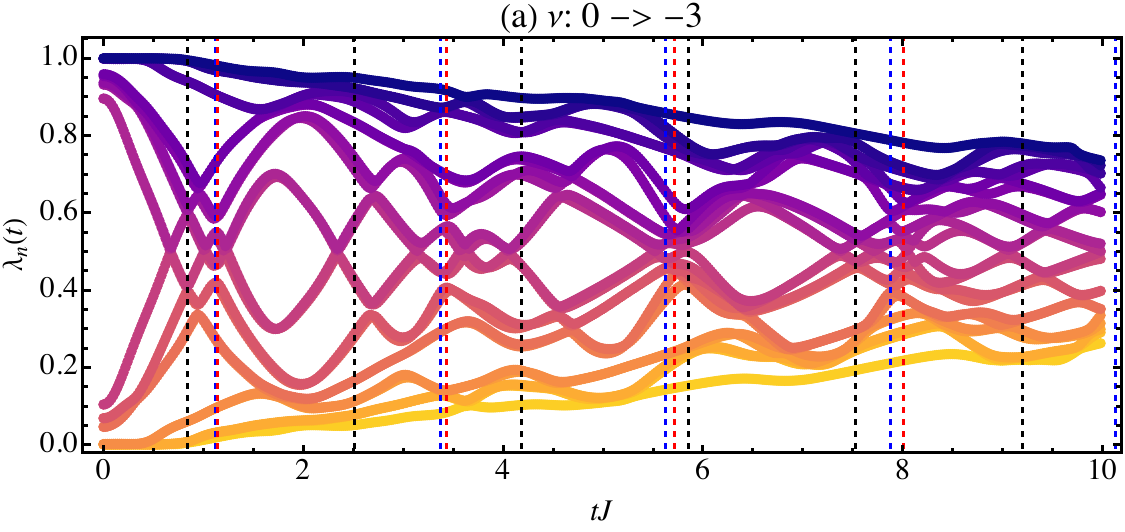}
\includegraphics[width=0.95\columnwidth]{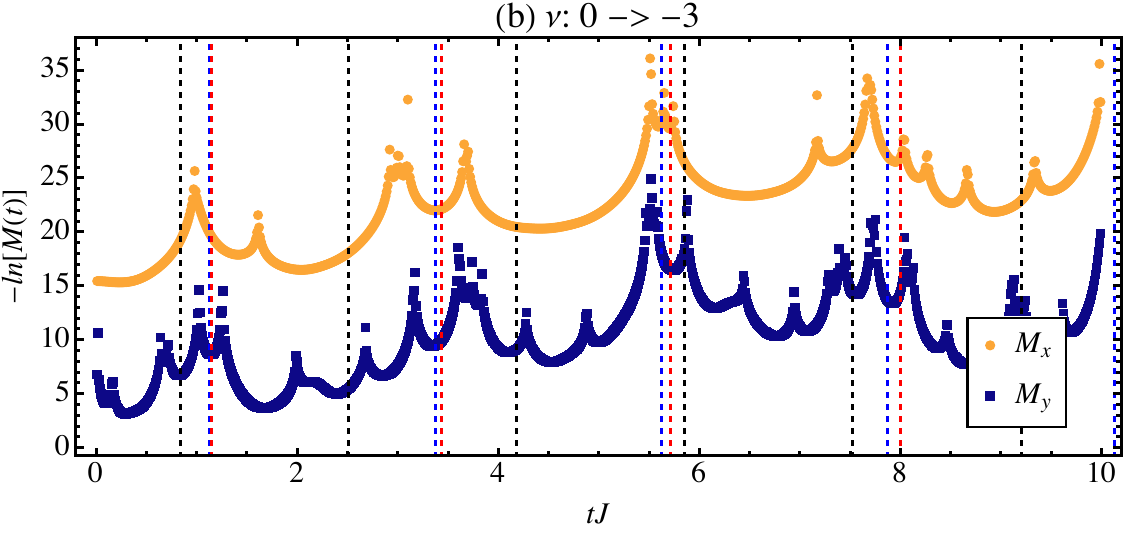}\\
\includegraphics[width=0.95\columnwidth]{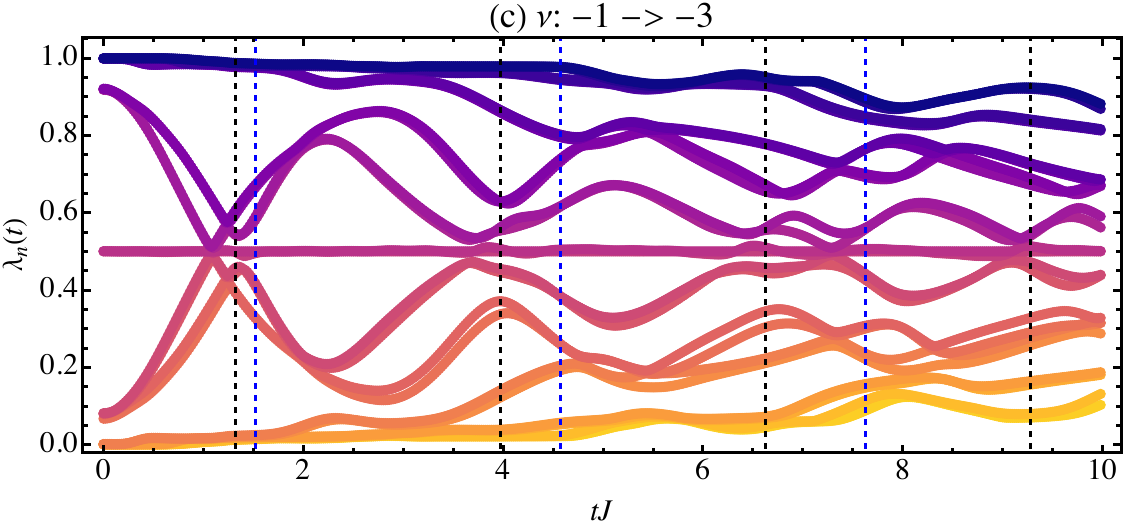}
\includegraphics[width=0.95\columnwidth]{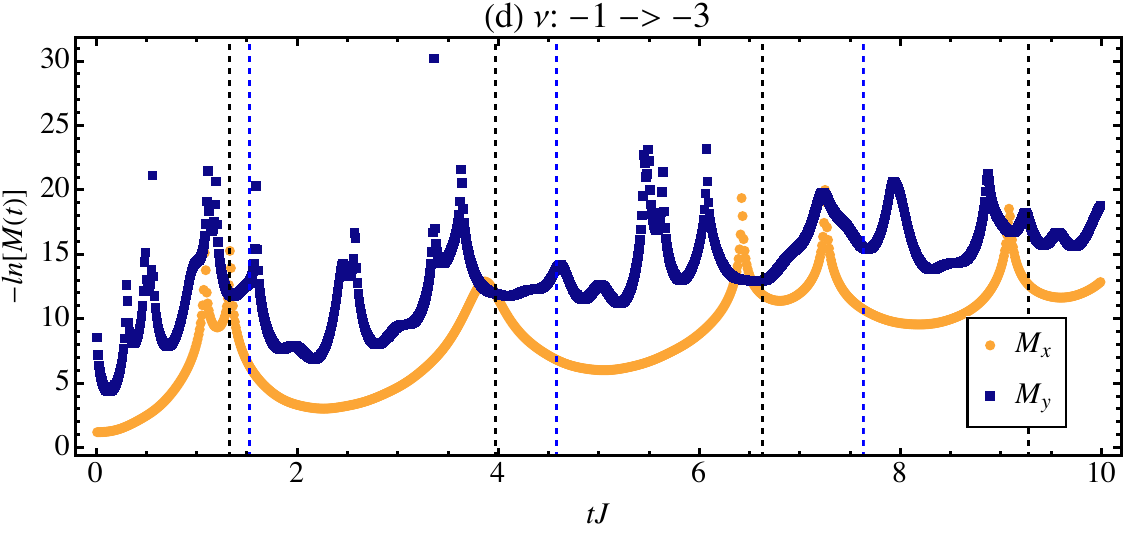}
\caption{Plots of $\mathcal{M}_{x,y}(t)$ and $\lambda_n(t)$ for different types of quenches, as marked on the figures, for the $R=3$ extended Kitaev chain. The system size is $N=350$, see Sec.~\ref{sec:kitmodel} for details of the parameters. For $\lambda_n(t)$ only the 22 eigenvalues closest to $0.5$ are plotted.  Vertical dashed lines show the DQPT critical times, with different colors referring to distinct critical momenta.}
\label{lrkc-t2}
\end{figure*}

The previous results demonstrate that the string order parameter can have non-trivial dynamics following a quench even in cases where the quench is between phases for which the string order parameter is zero. We therefore now widen our considerations to the $R=3$ extended Kitaev chain and topological phases with winding numbers $\nu\in\{0,-1,-2,-3\}$. Two examples of such quenches are shown in Fig.~\ref{lrkc-t2} for both $\lambda_n(t)$ and $\mathcal{M}_{x,y}(t)$. There are multiple critical times of the string order parameters, as well as many crossings at $0.5$ for $\lambda_n(t)$. None of the critical times coincide, and it is in general difficult to see any pattern in the resulting dynamics for the different quantities. Further examples can be found in appendix \ref{app:scaling} and more data for a variety of quenches in Ref.~\onlinecite{Bhattacharyya2026}. We can note also here that for $\nu:-1\to-3$ in Fig.~\ref{lrkc-t2}(c,d) the first critical times of $\mathcal{M}_x(t)$ and $\lambda_n(t)$ do coincide, but this is not repeated and appears to not be generic. See appendix \ref{app:scaling} for further discussion and details.

\section{Discussion and Conclusions}\label{sec:conc}

In conclusion we have investigated three different forms of dynamical criticality. The first are the zeroes of string order parameters, the second is the crossing of the dynamical entanglement spectrum $\lambda_n(t^*)=0.5$, and finally the critical times when the return rate becomes non-analytic, \emph{i.e.}~DQPTs. In all cases the dynamics is induced by a quench and we investigated the dependence of these critical times and general dynamics for quenches between a range of topological phases for two exemplary models. The main conclusion we come to is that, despite previous reports in the literature, there is no generic connection between these forms of criticality either quantitatively or qualitatively. Each can exist with or without the other and at different critical times. There is a general connection between the time scales, but this should not be surprising, and we interpret this in terms of the reshuffling of correlations in the systems following the quenches. In specific circumstances the string order and entanglement critical times can coincide, but this appears to be far from generic behavior, and indeed sometimes is actually for opposite quenches.

In general we find that different dynamical quantities do not serve as useful proxies for each other, and dynamical criticality is not a generic property but occurs at different times depending on the quantity being studied. This further suggests that there is no genuine sense to a dynamical phase or a DQPT as distinguishing macroscopic quantities during the dynamics in such cases. Indeed the string order parameters studied do not distinguish quenches between different possible topological phases.

In this work we have considered only the most elementary generalization, considering a wider range of topological phases but still of a simple two band model. Other interesting developments would be to consider multi-band models or terms which break the chiral symmetry. Indeed other topological classes from the topological periodic table could be considered. However finding appropriate string order parameters remains a difficult task. A further question would be the robustness of the results to disorder and interactions, particularly for the sensitive string order parameters. Additionally the behavior of the string order parameters away from the sudden quench limit would also be interesting to follow up.

\acknowledgments
This work was supported by the National Science Centre (NCN, Poland) by the grant 2024/53/B/ST3/02600 (NS, SG). Data for the numerical results, along with a larger selection of quenches, can be found at Ref.~\onlinecite{Bhattacharyya2026}.

\appendix

\section{Details on the SSH Model}\label{app:ssh}

\noindent
For the SSH model the Hamiltonian $\mathcal{H}_k$, one may note that Eq.~\eqref{ssh-hk}, is diagonalised by a unitary transformation $U^{\dagger}_k\mathcal{H}_k U_k$ where
\begin{equation}
U_k=\frac{1}{\sqrt{2}}
\begin{pmatrix}
e^{i\phi_k} & -e^{i\phi_k} \\
1 & 1
\end{pmatrix}.
\label{ssh-U} 
\end{equation}
If $\Phi_k=U_k^{\dagger}\Gamma_k$ is the diagonal momentum space basis then we can rewrite Eq.~\eqref{ssh2} as
\begin{equation}
H = \sum\limits_{k} \Phi_k^{\dagger} \begin{pmatrix}
\varepsilon_k & 0 \\
0 & -\varepsilon_k
\end{pmatrix} \Phi_k\,.
\end{equation}
For the initial Hamiltonian we define
\begin{equation}
\Phi_k^i = \frac{1}{\sqrt{2}}\begin{pmatrix}
e^{i\phi^i_k} a_k + b_k \\
e^{-i\phi^i_k} a_k + b_k
\end{pmatrix}\,,
\label{phi_i}
\end{equation}
and for the post-quench Hamiltonian we define
\begin{equation}
\Phi_k^f = \frac{1}{\sqrt{2}}\begin{pmatrix}
e^{i\phi^f_k} a_k + b_k \\
e^{-i\phi^f_k} a_k + b_k
\end{pmatrix}\,.
\label{phi_f}
\end{equation}
The relation between them is then
\begin{equation}
\Phi_k^f =U_f^{\dagger} U_i \Phi_k^i = \frac{1}{2}\begin{pmatrix}
T_1 & T_2 \\
T_2 & T_1
\end{pmatrix} \Phi_k^i\,,
\end{equation}
where $T_{1,2} = (1 \pm e^{-i\delta\phi_k})$ and we have also defined $\delta\phi_k = \phi^f_k - \phi^i_k$ for convenience. One then finds
\begin{align}
B^-_l(t)
=& \frac{1}{\sqrt{8N}} \sum\limits_k \left[ e^{ikl}e^{iE_k^f t}\left( T_1^{\dagger} \gamma_1^{\dagger} + T_2^{\dagger} \gamma_2^{\dagger} \right) 
\right. \nonumber \\&\left.
+ e^{ikl} e^{-iE_k^f t} \left( T_2^{\dagger} \gamma_1^{\dagger} + T_1^{\dagger} \gamma_2^{\dagger} \right) \right] - \textrm{H.c.}\,,
\end{align}
and similarly
\begin{align}
A^+_m(t)
=& \frac{1}{\sqrt{8N}} \sum\limits_k e^{-i\phi^f_k} \left[ e^{ikm}e^{iE_k^f t}\left( T_1^{\dagger} \gamma_1^{\dagger} + T_2^{\dagger} \gamma_2^{\dagger} \right) 
\right. \nonumber \\&\left.
- e^{ikm} e^{-iE_k^f t} \left( T_2^{\dagger} \gamma_1^{\dagger} + T_1^{\dagger} \gamma_2^{\dagger} \right) \right] + \textrm{H.c.}\,, \nonumber \\
\end{align}
for the operators which enter the string order parameter.

Here we also give the explicit expressions for the correlation functions which are needed to calculate the string order parameters. First we have $\langle A^+_l B^-_m \rangle_0=-\langle B^-_l A^+_m \rangle_0$ and
\begin{widetext}
\begin{align}
\langle B^-_l A^+_m \rangle_0 =
& \frac{1}{8N}\sum\limits_k e^{i\phi^f_k} e^{ik(l-m)}\left[ T_2^{\dagger}T_2 - e^{2iE_k^f t} T_2^{\dagger}T_1 + e^{-2iE_k^f t} T_1^{\dagger}T_2 - T_1^{\dagger}T_1 \right] \nonumber \\
&\qquad\qquad+ e^{-i\phi^f_k} e^{-ik(l-m)}\left[ T_2T_2^{\dagger} + e^{-2iE_k^f t} T_1T_2^{\dagger} - e^{2iE_k^f t} T_2T_1^{\dagger} - T_1T_1^{\dagger} \right] \nonumber \\
=& \frac{1}{N}\sum\limits_k \left[ -\cos(\delta\phi_k)\cos[\phi^f_k + k(l-m) ] - \sin(\delta\phi_k)\sin[\phi^f_k + k(l-m) ] \cos(2E_k^f t) \right]\,.
\label{CB}
\end{align}
Secondly we also have $\langle A^+_l A^+_m \rangle_0=-\langle B^-_l B^-_m \rangle_0^*$ and
\begin{align}
\langle A^+_l A^+_m \rangle_0
=& \frac{1}{8N}\sum\limits_k e^{ik(l-m)}\left[ T_2^{\dagger}T_2 - e^{2iE_k^f t} T_2^{\dagger}T_1 - e^{-2iE_k^f t} T_1^{\dagger}T_2 + T_1^{\dagger}T_1 \right] 
\nonumber \\
&\qquad\qquad+e^{-ik(l-m)}\left[ T_2T_2^{\dagger} - e^{-2iE_k^f t} T_1T_2^{\dagger} - e^{2iE_k^f t} T_2T_1^{\dagger} + T_1T_1^{\dagger} \right] \nonumber \\
=& \frac{1}{N}\sum\limits_k \left[ \cos k(l-m) +-i\sin(\delta\phi_k)\sin k(l-m) \sin(2E_k^f t) \right]\,.
\label{BB}
\end{align}
\end{widetext}
These are the inputs for the string order parameter determinant.

\section{Details on the Extended Kitaev Model}\label{app:kit}
\begin{figure}
\includegraphics[width=0.95\columnwidth]{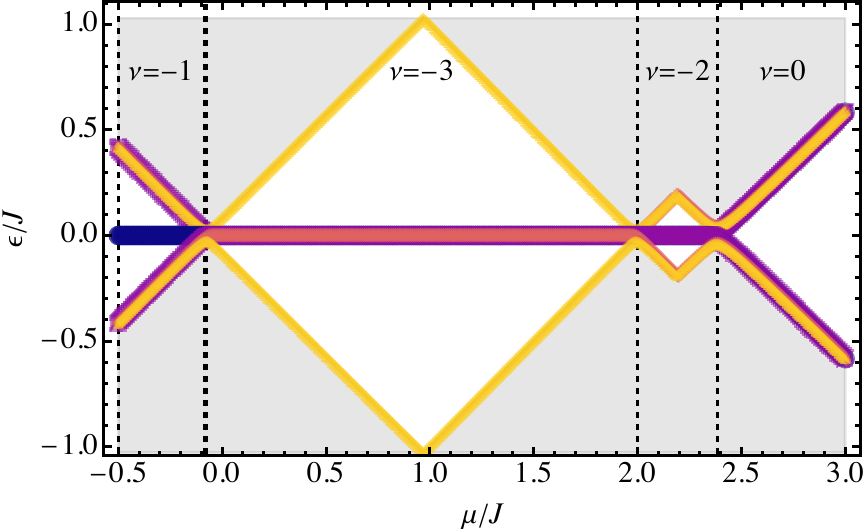}
\caption{Plots of the spectrum as a function of chemical potential $\mu$ for the extended Kitaev model, compare with Fig.~\ref{fig:kiteq}. The hopping is $\vec{J}=(1,1,1)J$ and pairing is $\vec\Delta=(1.2,1.4,1.6)J$ with $R=3$. Gray shows the bulk spectrum and the smallest eight eigenvalues of the open system are shown in color. Phase boundaries are given by dashed black lines.}
\label{fig:kitsp}
\end{figure}

\noindent
For the extended Kitaev chain the Hamiltonian $\mathcal{H}_k$, Eq.~\eqref{ekc-hk}, is diagonalised by the unitary transformation $U^{\dagger}_k\mathcal{H}_k U_k$ where
\begin{equation}
U_k=
\begin{pmatrix}
\cos{\frac{\theta_k}{2}} & i\sin{\frac{\theta_k}{2}} \\
i\sin{\frac{\theta_k}{2}} & \cos{\frac{\theta_k}{2}}
\end{pmatrix}.
\label{kit-U} 
\end{equation}
with $\tan\theta_k = d^y_k / d^z_k$. This results in 
\begin{equation}
H=\sum\limits_k \Phi_k^{\dagger} {U}^{\dagger}_k \mathcal{H}_k {U}_k \Phi_k 
=\sum\limits_k \Phi_k^{\dagger} \begin{pmatrix}
\varepsilon_k & 0 \\
0 & -\varepsilon_k
\end{pmatrix} \Phi_k
\end{equation}
where ${\Phi}_k={U}_k^{\dagger}{\Psi}_k$. See Fig.~\ref{fig:kitsp} for an example of the spectra showing the zero modes in different topological phases as a function of $\mu$ to compare with Fig.~\ref{fig:kiteq}.

Using $\Phi_k={U}_k^{\dagger}{\Psi}_k$ we can define the bases of the initial and post-quench Hamiltonians respectively:
\begin{equation}
\Phi_k^i = \begin{pmatrix}
\cos{\frac{\theta^i_k}{2}} c_k - i\sin{\frac{\theta^i_k}{2}} c^{\dagger}_{-k} \\\;
- i\sin{\frac{\theta^i_k}{2}} c_k + \cos{\frac{\theta^i_k}{2}} c^{\dagger}_{-k}
\end{pmatrix}\,,
\label{ekc-phi_i}
\end{equation}
and
\begin{equation}
\Phi_k^f = \begin{pmatrix}
\cos{\frac{\theta^f_k}{2}} c_k - i\sin{\frac{\theta^f_k}{2}} c^{\dagger}_{-k} \\\;
- i\sin{\frac{\theta^f_k}{2}} c_k + \cos{\frac{\theta^f_k}{2}} c^{\dagger}_{-k}
\end{pmatrix}\,.
\label{ekc-phi_f}
\end{equation}
Combining Eq.~\eqref{ekc-phi_i} in Eq.~\eqref{ekc-phi_f} we find
\begin{equation}
\Phi_k^f = U_f^{\dagger} U_i \Phi_k^i 
= \frac{1}{2}\begin{pmatrix}
\cos(\delta\theta_k) & -i\sin(\delta\theta_k)\\
-i\sin(\delta\theta_k) & \cos(\delta\theta_k)
\end{pmatrix} \Phi_k^i\,,
\end{equation}
where $\delta\theta_k = (\theta^f_k - \theta^i_k)/2$.

For the extended Kitaev chain we also need the following correlation functions for \eqref{kitfm} which follow easily from the above:
\begin{widetext}
\begin{align}
\langle \Phi^{-}_{m}(t)\Phi^{+}_{n}(t)\rangle_0
=& \dfrac{2}{N} \sum\limits_{k>0} \left[ -\cos (2\delta\theta_k) \cos \left[ \theta^f_k - k(m-n) \right] - \sin(2\delta\theta_k)\sin \left[ \theta^f_k - k(m-n) \right] \cos2\varepsilon_{k}^{f}t \right]\,,\\
\langle \Phi^{+}_{m}(t)\Phi^{-}_{n}(t)\rangle_0
=& \dfrac{2}{N} \sum\limits_{k>0} \left[ \cos(2\delta\theta_k) \cos \left[\theta^f_k + k(m-n) \right] + \sin(2\delta\theta_k) \sin \left[ \theta^f_k +k(m-n) \right] \cos2\varepsilon_{k}^{f}t \right]\,,\\
\langle \Phi^{-}_{m}(t)\Phi^{-}_{n}(t)\rangle_0
=& \dfrac{2}{N} \sum\limits_{k>0} \left[ -\cos k(a-b) + i \sin(2\delta\theta_k) \sin [k(m-n)] \sin2\varepsilon_{k}^{f}t \right]\,,\textrm{ and}\\
\langle \Phi^{+}_{m}(t)\Phi^{+}_{n}(t)\rangle_0
=& \dfrac{2}{N} \sum\limits_{k>0} \left[ \cos [k(m-n)] + i \sin(2\delta\theta_k) \sin [k(m-n)]) \sin2\varepsilon_{k}^{f}t \right]\,.
\label{lrkc1-pp}
\end{align}
\end{widetext}

\section{Additional Results}\label{app:scaling}

\begin{figure}
\includegraphics[width=0.95\columnwidth]{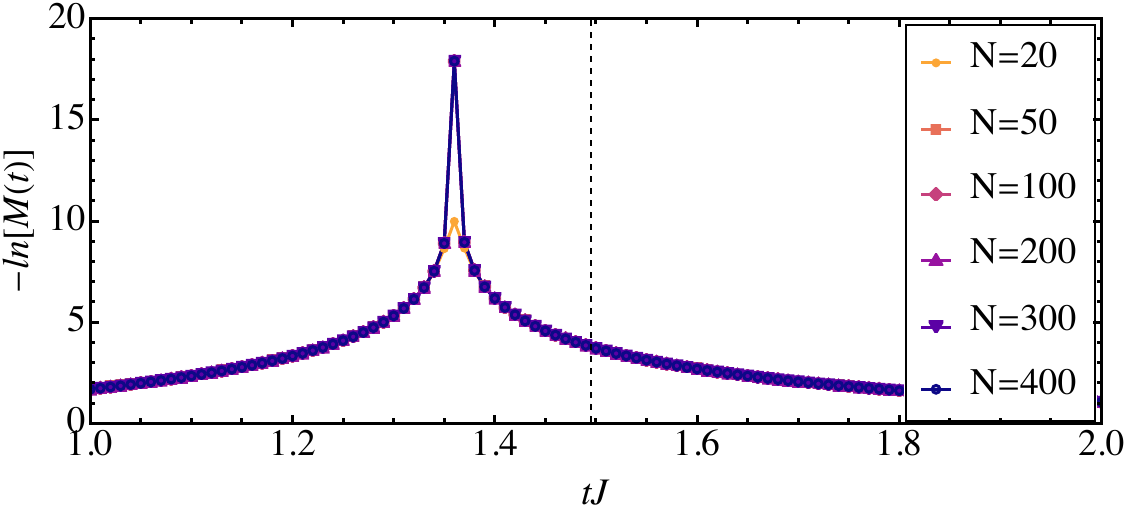}
\caption{$\mathcal{M}(t)$ for the quench in the SSH model $\nu:1\to0$ with $|\delta|=0.4$, plotted near the first  DQPT critical time marked by the dashed line. Results for a range of system sizes are shown. As can be seen the string order parameter is well converged even for $N=50$ and becomes zero at $t'=1.36/J$. By contrast $t_c=1.495/J$.}
\label{fig:app-ssh-sop-sc}
\end{figure}

\begin{figure}
\includegraphics[width=0.95\columnwidth]{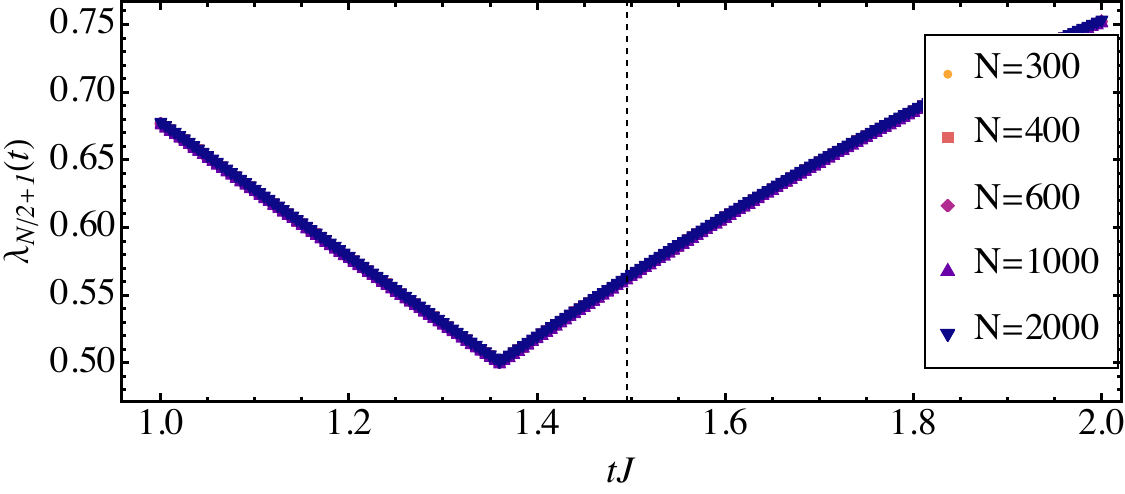}
\caption{The central entanglement eigenvalue for the quench in the SSH model $\nu:0\to1$ with $|\delta|=0.4$, plotted near the first crossing at $\lambda_{N/2,N/2+1}(t^*)=0.5$. The vertical dashed line is the DQPT critical time. Results for a range of system sizes are shown. As can be seen the entanglement eigenvalues are well converged even for $N=50$ and cross at $t^*=1.36/J$. By contrast $t_c=1.495/J$.}
\label{fig:app-ssh-es-sc}
\end{figure}

In this appendix we present several more results of the string order parameter and entanglement spectra, including finite size scaling. We note that in general the DQPT critical time is known to arbitrary accuracy in the thermodynamic limit. It is more difficult to extract the necessary information from the correlation matrix~\cite{Jones2019}. As we will give examples for here, we do not see any shift of the SOP and entanglement critical times over system sizes we can check. In Fig.~\ref{fig:app-ssh-sop-sc} we check the scaling of the string order parameter zero near the first DQPT critical time and in Fig.~\ref{fig:app-ssh-es-sc} we check the scaling of the first crossing of $\lambda_n(t)$ at $0.5$. These have already converged even for $N=50$ and shows no change up to $N=2000$, in the case of the entanglement. The $\lambda_n(t)$ crossing is consistently before the DQPT critical time $t_c=1.495/J$ at $t^*=1.36/J$, and the string order parameter zero also at $t'=1.36/J$, though we must note that these are in fact for opposite quenches. This is a curious fact but is only applicable to the SSH model due to the symmetry of the bulk for $\nu\in\{0,1\}$ with the same magnitude of $\delta$ but opposite signs. No such phenomena can be seen in any other case. Nonetheless it requires explanation even in this special case. For the quench $\nu:0\to1$ the $\lambda_n(t)$ crossing can be thought of as the system mimicking the topological entanglement at this time. For the opposite quench $\mathcal{M}(t)=0$ means conversely that the string order parameter is suggesting the dynamical system is momentarily in a simple trivial phase. These occur at the same critical times due to the aforementioned symmetry.

\begin{figure*}
\includegraphics[width=0.95\columnwidth]{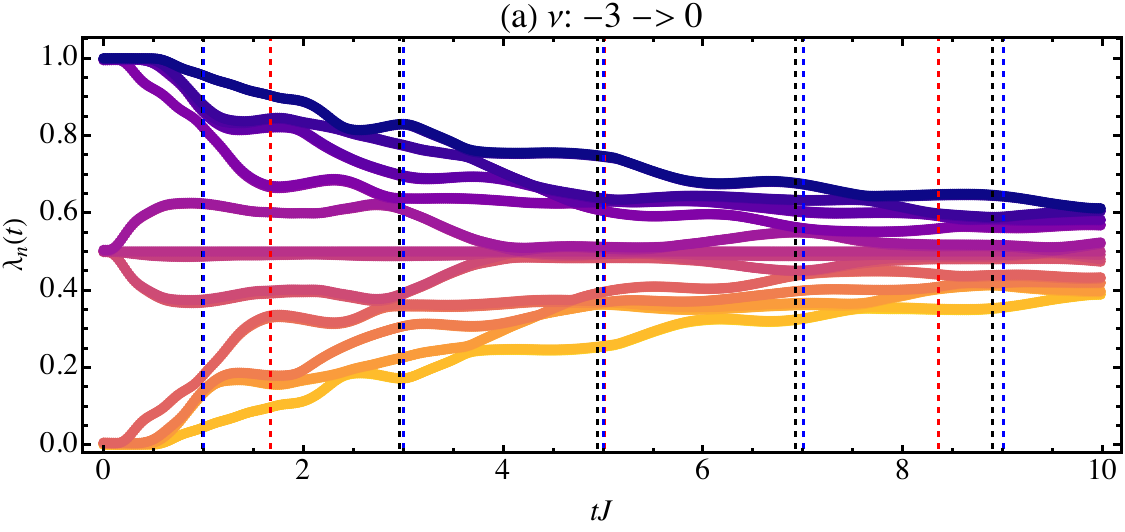}
\includegraphics[width=0.95\columnwidth]{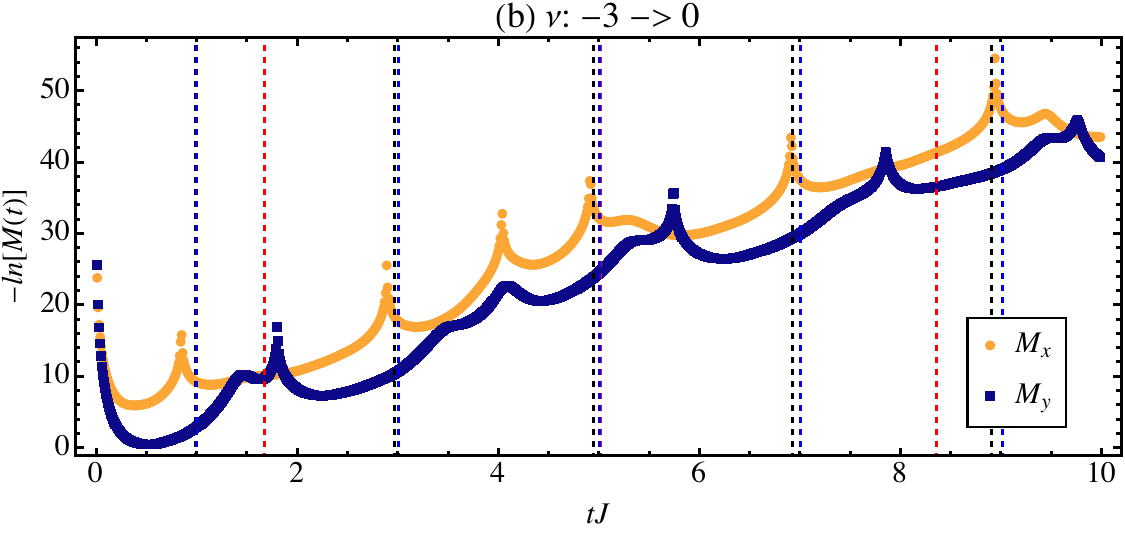}\\
\includegraphics[width=0.95\columnwidth]{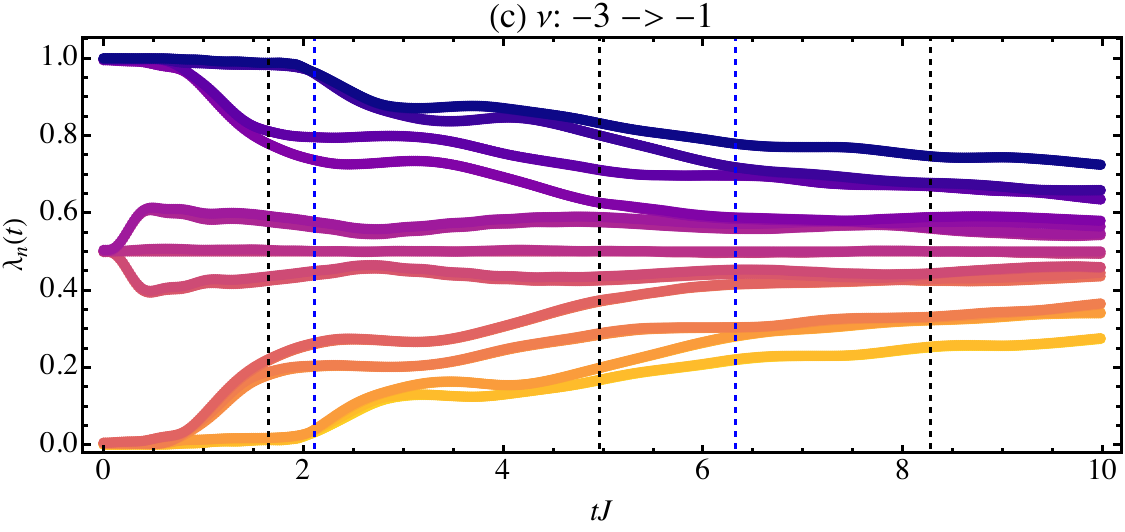}
\includegraphics[width=0.95\columnwidth]{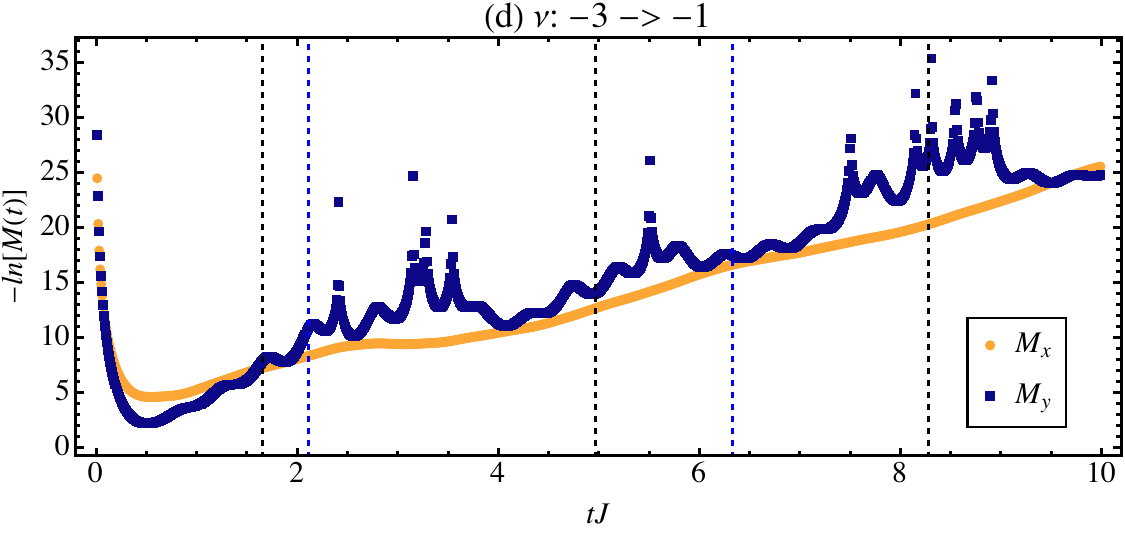}
\caption{Plots of $\mathcal{M}_{x,y}(t)$ and $\lambda_n(t)$ for different types of quenches, as marked on the figures, for the $R=3$ extended Kitaev chain. The system size is $N=350$, see Sec.~\ref{sec:kitmodel} for details of the parameters. For $\lambda_n(t)$ only the 22 eigenvalues closest to $0.5$ are plotted.  Vertical dashed lines show the DQPT critical times, with different colors referring to distinct critical momenta.}
\label{fig:app-lrkc-t}
\end{figure*}

Fig.~\ref{fig:app-lrkc-t} shows several more examples of quenches for the $R=3$ extended Kitaev chain. For such quenches there is little apparent connection between the three phenomena being investigated: DQPTs, string order parameter zeroes, and entanglement spectra crossings.

\begin{figure}
\includegraphics[width=0.95\columnwidth]{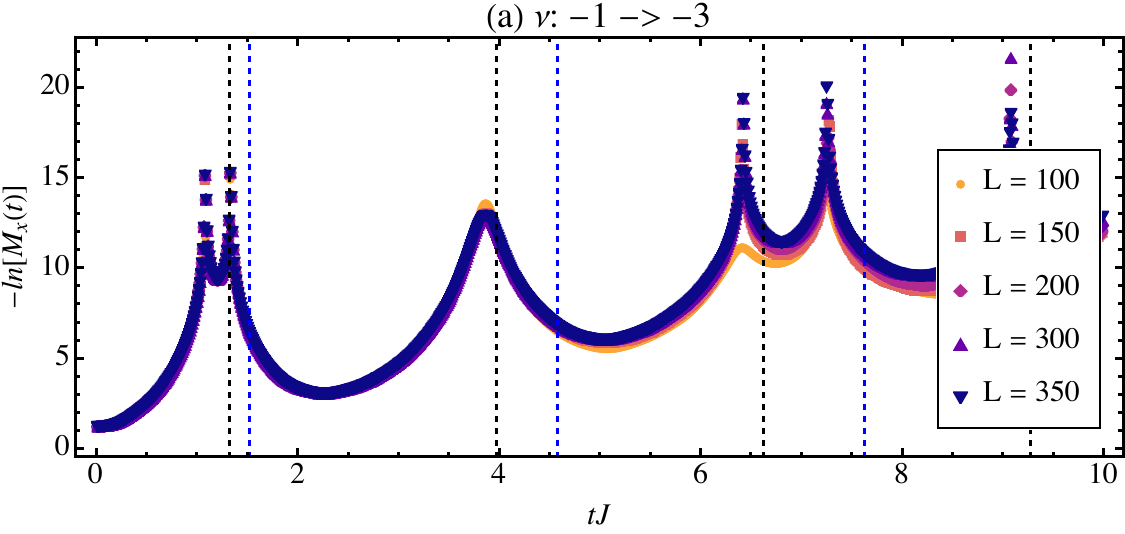}\\
\includegraphics[width=0.95\columnwidth]{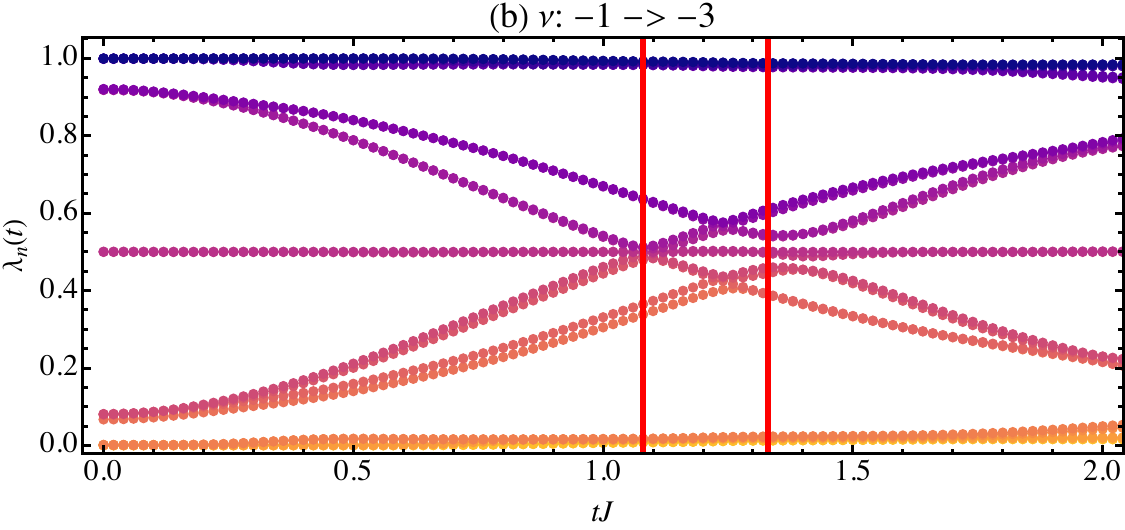}
\caption{(a) Finite size scaling of $\mathcal{M}_x(t)$ for the $R=3$ extended Kitaev chain. (b) The central entanglement eigenvalues for the same quench as (a). The vertical lines are the location of DQPT critical times.}
\label{fig:app-kit-sc}
\end{figure}

In Fig.~\ref{fig:app-kit-sc}(a) finite size scaling of $\mathcal{M}_x(t)$ is shown for the quench $\nu:-1\to-3$ for the $R=3$ extended Kitaev chain. As can be seen the zeroes are already converged. We extract the locations of the zeroes and compare them to the entanglement spectrum crossings. For this particular case the first critical times for the entanglement spectrum crossing and the string order parameter coincide, but do not coincide with the DQPT critical time, but we stress once again this is not generic behavior.

\section{The Topological Index}\label{app:top}

For a two band model with a chiral symmetry $\mathcal{S}=\sigma^x$, $\mathcal{S}H_k\mathcal{S}^{-1}=-H_k$ and $\mathcal{S}^2=1$, one can write $H_k=\vec{d}_k\cdot\vec{\sigma}$ with $\vec{d}_k=(0,d^y_k,d^z_k)$. The appropriate topological invariant is then the chiral invariant
\begin{equation}
    \nu=\frac{1}{4\pi i}\int_{-\pi}^\pi dk\textrm{tr}\,\mathcal{S}H_k\partial_kH_k^{-1}
\end{equation}
and one can find
\begin{equation}
    \nu=-\frac{2}{\pi}\int_{-\pi}^\pi dk\frac{d^x_k\partial_kd^z_k-d^z_k\partial_kd^x_k}{(d^x_k)^2+(d^z_k)^2}.
\end{equation}
The models we consider also have particle-hole symmetry $\mathcal{P}=\sigma^x\mathcal{K}$, $\mathcal{P}H_k\mathcal{P}^{-1}=-H_{-k}$ and $\mathcal{P}^2=1$ where $\mathcal{K}$ is the conjugation operator. And we have a time reversal symmetry $\mathcal{T}=\mathcal{K}$, $\mathcal{T}H_k\mathcal{T}^{-1}=H_{-k}$ where $\mathcal{T}^2=1$. Note that the chiral symmetry is $\mathcal{S}=\mathcal{P}\mathcal{T}$. This means we are in the BDI class~\cite{Kitaev2009,Ryu2010,Teo2010}.


%

\end{document}